\documentclass{JHEP3}

\usepackage{amssymb}
\usepackage{amsfonts}
\usepackage{amsbsy}
\usepackage{amsmath}
\usepackage{amsthm}
\usepackage{graphicx}
\usepackage{epstopdf}
\usepackage{multirow}
\usepackage{latexsym}
\usepackage{array}

\input cyracc.def 
\newfont{\twelvecyr}{wncyr10 at 12pt}

\newcommand{\new}[1]{{\em #1}}

\def\Z{\mathbb{Z}}
\def\F{\mathbb{F}}
\def\Q{\mathbb{Q}}
\def\C{\mathbb{C}}

\def\G{\mathfrak{g}}

\def\P{\mathbb{P}}

\def\n3a{t}

\def\tf{\tilde{\F}}

\def\ge{{\mathfrak{e}}}
\def\gso{{\mathfrak{so}}}
\def\gsu{{\mathfrak{su}}}

\def\gf{{\mathfrak{f}}}
\def\gg{{\mathfrak{g}}}

\def\ho{h^{1, 1}}
\newcommand{\eq}[1]{(\ref{#1})}

\title{Toric bases for 6D F-theory models}

\author{David R.  Morrison$^{1}$ and Washington Taylor$^2$\\
$^1$Departments of Mathematics and  
Physics\\ University of California, Santa Barbara\\ Santa Barbara, CA 93106, USA\\
\\
$^2$Center for Theoretical Physics\\
Department of Physics\\
Massachusetts Institute of Technology\\
77 Massachusetts Avenue\\
Cambridge, MA 02139, USA\\
\\
{\tt drm} {\rm at} {\tt math.ucsb.edu},
{\tt wati} {\rm at} {\tt mit.edu}
}

\preprint{UCSB Math 2012-14, MIT-CTP-4356}

\abstract{We find all smooth toric bases that support elliptically
  fibered Calabi-Yau threefolds, using the intersection structure of
  the irreducible effective divisors on the base.  These bases can be
  used for F-theory constructions of six-dimensional quantum
  supergravity theories.  There are 61,539 distinct possible toric
  bases.  The associated 6D supergravity theories have a number of
  tensor multiplets ranging from 0 to 193.  For each base an explicit
  Weierstrass parameterization can be determined in terms of the toric
  data.  The toric counting of parameters matches with the
  gravitational anomaly constraint on massless fields.  For bases
  associated with theories having a large number of tensor multiplets,
  there is a large non-Higgsable gauge group containing multiple
  irreducible gauge group factors, particularly those having algebras
  $\ge_8, \gf_4,$ and $\gg_2 \oplus \gsu(2)$ with minimal
  (non-Higgsable) matter.}

\begin{document}

\section{Introduction}

F-theory \cite{Vafa-f, Morrison-Vafa, Morrison-Vafa-II} provides a
very general and nonperturbative approach to constructing vacua of
string theory.  In six dimensions, there is a large connected moduli
space of F-theory vacua.  Branches of this moduli space are described
by F-theory on elliptic fibrations over different bases $B$ that are
complex surfaces.  Branches associated with different bases are
connected through tensionless string transitions \cite{dmw,
  Seiberg-Witten, Morrison-Vafa-II}.  Over each specific base surface
$B$ there is a rich space of physical theories with different gauge
groups
and
matter content, characterized by different configurations of 7-branes
on $B$ associated with different singularity structures in the
elliptic fibration over $B$.  The geometry of an F-theory model is
closely mirrored in the structure of the resulting supergravity
theory.  For example, the number of tensor multiplets $T$ in the 6D
supergravity theory from an F-theory compactification corresponds to
the topological structure of $B$ through $T = h^{1, 1} (B) -1$.  6D
supergravity theories with a general number of tensor multiplets that
satisfy anomaly cancellation were originally described in
\cite{Romans, Sagnotti}.  Much of the study of 6D F-theory models has
focused on a simple set of base manifolds with small $T$.  In
particular, the Hirzebruch surfaces $\F_m$ give F-theory models with
$T = 1$ that are dual to heterotic compactifications on a smooth K3
\cite{Morrison-Vafa-II}.  Recent work \cite{0,
  Morrison-Taylor, Braun} has explored the set of models over the
simplest F-theory base $\P^2$, with $T = 0$.  In this paper we explore
the space of all F-theory bases by explicitly constructing and
enumerating all smooth surfaces that have a description in terms of
toric geometry and can act as bases for an elliptically-fibered
Calabi-Yau threefold.

The approach we take here to describing bases for elliptic fibrations is
based on the intersection structure of divisors on the base.  In a
general F-theory model, the gauge group and matter content of the
corresponding supergravity theory are determined by the structure of
the elliptic fibration.  This structure is encoded physically in the
positions of 7-branes on $B$.  Elliptic fibrations can be described
through a Weierstrass model $y^2 = x^3 + fx + g$, where $f, g$ are
sections of certain line bundles over the base $B$.  By varying the
continuous parameters in $f$ and $g$ that describe the fibration, the
7-branes can be moved to increase or decrease the gauge group and
massless matter content, corresponding to Higgsing and unHiggsing
fields in the supergravity theory.  For many bases, such as the
Hirzebruch surfaces $\F_m, m > 2$, the Weierstrass coefficients $f$
and $g$ must vanish on certain codimension one cycles (divisors) on
the base $B$.  This gives rise to a certain gauge group content for
each base that cannot be removed through Higgsing by massless matter
fields in the theory.  In a recent paper \cite{clusters}, we gave a
systematic analysis of the intersection structure of divisors for any
base.  We identified a set of irreducible blocks (``non-Higgsable
clusters'') composed of one or more intersecting divisors of
self-intersection -2 or less that impose a non-Higgsable gauge group,
sometimes combined with a fixed massless matter representation, on the
resulting supergravity theory.  In this paper we use this set of
blocks as a guide in the construction of all toric bases.  For toric
bases, the configuration of irreducible effective divisors with
negative self-intersection is very simple, and always lies within a
closed linear chain, greatly simplifying the analysis.  A challenge
for future work is to use related methods to analyze more general,
non-toric bases.  Previous work has been done using toric geometry in
another way to classify elliptically fibered threefolds for F-theory
compactifications \cite{Candelas-pr, Braun}.  In that work, toric methods
were used to construct the entire threefold for the compactification
as a hypersurface in a toric variety (compact Calabi-Yau manifolds
cannot themselves be toric).  The approach taken here simplifies the
story by focusing on the geometry of the base.

For a given toric base, the methods of toric geometry give a simple
classification of the monomials that appear in the Weierstrass
coefficients $f, g$.  This gives an explicit Weierstrass
parameterization for every toric base.  We clarify some subtleties in
the counting of degrees of freedom and confirm that the number of
degrees of freedom in the Weierstrass models matches with the number
of massless scalar fields $H$ expected from the gravitational anomaly
condition $H-V= 273-29T$, where $V$ is the number of vector
multiplets.  This explicit parameterization can be used for further
analysis of the space of theories over any given base.  It also
confirms for any given base that the minimal gauge and matter content
is precisely that expected from the non-Higgsable clusters in the set
of irreducible divisors.

In Section \ref{sec:basics} we review some basic aspects of F-theory,
algebraic geometry, and toric geometry that will be useful.  In
Section \ref{sec:toric} we describe the complete set of toric bases
for 6D F-theory models.  In Section \ref{sec:Weierstrass} we show how
an explicit Weierstrass parameterization can be easily constructed for
any toric F-theory base, and describe how the number of degrees of freedom
in the theory matches the  toric data.
Section \ref{sec:further}  suggests natural extensions of this
analysis to non-toric bases and four-dimensional F-theory models, and
Section \ref{sec:conclusions} contains concluding remarks.

\section{Geometry of bases for 6D F-theory models}
\label{sec:basics}

In this section we briefly review some  aspects of F-theory
and geometry that will be useful in this work.
Some of the basic aspects of the ideas used here are described in more
detail in \cite{clusters}.  Further pedagogical introductions to
F-theory can be found in \cite{Morrison-TASI, Denef-F-theory, WT-TASI}.

\subsection{Base surfaces for F-theory models}

A six-dimensional F-theory model is constructed from an elliptic fibration
with section over a base $B$, where the total space is a Calabi-Yau
threefold.  The elliptic fibration for a 6D F-theory model can be
described by a Weierstrass form
\begin{equation}
y^2 = x^3 + fx + g \,.
\label{eq:Weierstrass}
\end{equation}
Here $f, g$ and the discriminant locus
\begin{equation}
\Delta = 4 f^3 + 27 g^2
\label{eq:discriminant}
\end{equation}
are sections of line bundles
\begin{equation}
f \in -4K, \; \; g \in -6K, \; \; \Delta \in -12K
\end{equation}
where $K$ is the canonical class of $B$.  The intersection form on
$H_2 (B,\Z)$ controls much of the physics of the associated 6D
supergravity theory \cite{KMT-II, Seiberg-Taylor}.  This intersection
form has signature $(1, T)$, where $T =\ho (B) -1$ is the number of
tensor multiplets in the 6D supergravity theory.  The discriminant
locus $\Delta$ can be described physically in terms of 7-branes in the
language of type IIB string theory \cite{Schwarz}.
These 7-branes are wrapped on divisor classes where $\Delta$ vanishes.
Singularities of the elliptic fibration can occur where the
Weierstrass coefficients $f$ and $g$ vanish.  Such codimension one
singularities were classified by Kodaira \cite{Kodaira} and correspond
to a nonabelian gauge group in the 6D supergravity theory, the algebra of which can be computed using the Tate algorithm
\cite{Morrison-Vafa-II, Bershadsky-all, Morrison-sn,
  Grassi-Morrison-2}.  The type of singularity on a curve $D$ is
determined by the degrees of vanishing of $f, g, \Delta$ on that
curve.  For example, when $f$ and $g$ vanish to orders 4 and 5,
$\Delta$ vanishes to order 10 and the singularity gives rise to an
$\ge_8$ gauge algebra.  When $f, g$ vanish to order $\geq 4$ and $\geq 6$,
$\Delta$ vanishes to order 12 and the singularity cannot be resolved
in a fashion compatible with the Calabi-Yau condition on the total
space of the fibration.  Tables of singularities and symmetries are
found in many F-theory papers and reviews, including
\cite{Morrison-Vafa-II, Bershadsky-all, Grassi-Morrison-2,
  clusters}.

\begin{figure}
\begin{center}
\begin{picture}(200,130)(- 93,- 55)
\thicklines
\put(-175, 25){\line(1,0){50}}
\put(-150,32){\makebox(0,0){$-m\leq -3$}}
\put(-150,-33){\makebox(0,0){\small $\gsu(3), \gso(8), \gf_4$}}
\put(-150,-47){\makebox(0,0){\small $\ge_6, \ge_7, \ge_8$}}
\put(-70,55){\line(1,-1){40}}
\put(-30,35){\line(-1,-1){40}}
\put(-50,45){\makebox(0,0){-3}}
\put(-50,5){\makebox(0,0){-2}}
\put(-50,-40){\makebox(0,0){\small $\gg_2 \oplus \gsu(2)$}}
\put(30,70){\line(1,-1){40}}
\put(30,20){\line(1,-1){40}}
\put(70,45){\line(-1,-1){40}}
\put(45,65){\makebox(0,0){-3}}
\put(44,31){\makebox(0,0){-2}}
\put(60, 0){\makebox(0,0){-2}}
\put(50,-40){\makebox(0,0){\small $\gg_2 \oplus \gsu(2)$}}
\put(130,70){\line(1,-1){40}}
\put(130,20){\line(1,-1){40}}
\put(170,45){\line(-1,-1){40}}
\put(145,65){\makebox(0,0){-2}}
\put(144,31){\makebox(0,0){-3}}
\put(160, 0){\makebox(0,0){-2}}
\put(150,-40){\makebox(0,0){\small $\gsu(2) \oplus \gso(7) \oplus \gsu(2)$}}
\end{picture}
\end{center}
\caption[x]{\footnotesize    Clusters of intersecting
  curves that must carry a nonabelian gauge group factor.  For each
cluster the corresponding gauge algebra is noted and the gauge algebra and
matter content are listed in Table~\ref{t:clusters}}
\label{f:clusters}
\end{figure}
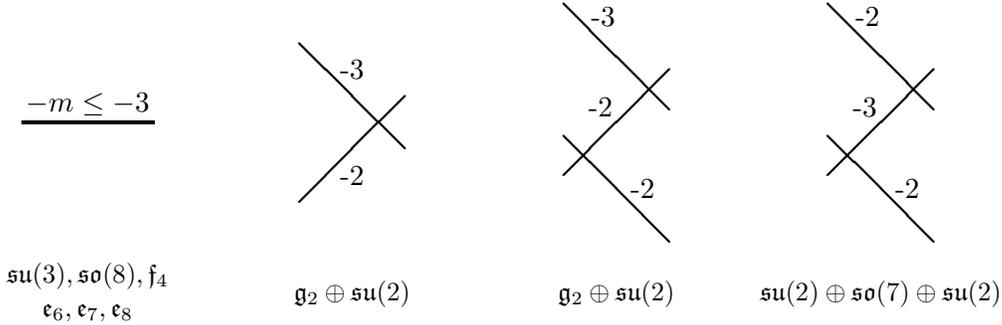

When $B$ contains an irreducible effective divisor $D$ of
self-intersection $-3$ or below, $-K \cdot D < 0,$ $f$ and $g$ must
vanish on $D$ to at least order 2, and a nonabelian gauge theory must
appear in the corresponding 6D supergravity theory.  In general this
situation is described by the ``Zariski decomposition'' 
of $A$ \cite{Zariski,cutkosky}, where
any effective divisor $A$ can be decomposed over the rational numbers as
\begin{equation}
A = \sum_{i}\zeta_i C_i + Y, \;\;\; \zeta_i \in\Q \,.
\label{eq:Zariski}
\end{equation}
where $Y$ (the ``free part'') satisfies $Y \cdot C \geq 0$ for all
curves $C$ on the surface and  $Y\cdot C_i=0$
for the irreducible effective $C_i$ such that $A\cdot C_i<0$,
and $\sum \zeta_i C_i$ (the ``fixed part'') contains all irreducible
divisors $C_i$ with $A\cdot C_i<0$.
For example, for a curve $D$ of self-intersection $-3$, the Zariski
decomposition of the anti-canonical divisor is $-K= (1/3)D + Y$, so 
sections $f, g$ of $-4K, -6K$ must vanish at least twice ($\geq 4/3,
\geq 2$ times) on $D$, and thus must carry a gauge group  the
algebra\footnote{Because the gauge group can have a quotient by a
  discrete subgroup, it is more precise to state this condition for
  the algebra, rather than for the group.}  of which contains
$\gsu(3)$.  Similarly,
a divisor $D$ of self-intersection $-4$ must carry an $\gso(8)$, and
for $D \cdot D < -4$ various other gauge algebras must appear, up to
curves of self-intersection $-12$ that carry an $\ge_8$ gauge algebra.
Furthermore, certain combinations of intersecting curves with negative
self-intersection must carry more complicated gauge group factors with
charged matter.  For example, a $-3$ curve intersecting a $-2$ curve
must carry a gauge algebra $\gg_2 \oplus \gsu(2)$ and charged matter
living in the $({\bf 7 + 1}, \frac{1}{2}{\bf 2})$ representation of
this algebra.  A complete analysis of all such possibilities is given
in \cite{clusters}; the results of this analysis are summarized in
 Table~\ref{t:clusters}, with the associated geometries depicted
in Figure~\ref{f:clusters}.  The clusters with a single curve of
negative self-intersection are all familiar from $\F_m$ bases, and the
gauge groups and matter content associated with multiple-curve
clusters were previously encountered in the field theory context in
\cite{Intriligator}.  In addition to the various ``non-Higgsable''
clusters that must carry a nontrivial gauge group factor and (in some cases)
matter, there can also be configurations of intersecting $-2$ curves.
$f$ and $g$ do not need to vanish on the $-2$ curves, so there is no
gauge group enforced on these blocks.  The gauge group over any
irreducible curve can have a larger algebra than the minimal algebra
indicated in Table~\ref{t:clusters}, if the Weierstrass coefficients
$f, g$ vanish at higher order then the minimal order required.  In
general, however, such larger gauge groups can be broken by Higgsing
massless matter fields, and for generic Weierstrass coefficients over
any given base the gauge algebra and matter content are expected to
have the minimal form.  We do not have a proof that this is true in
general, but, as described in Section \ref{sec:Weierstrass}, this
property holds for all the toric bases constructed in this
paper.

\begin{table}
\begin{center}
\begin{tabular}{| c | c | c | c  | c| c | 
c |
}
\hline
Diagram & Algebra & V & matter & $\zeta_i$ & $(f, g, \Delta)$ 
& $\Delta T_{\rm max}$
\\
\hline
$-3$ & $\gsu(3) $ & $8$ & 0&$1/3$& $(2, 2, 4)$ 
&   1/3
\\ 
\hline
$-4$ &$\gso(8) $ & $28$&0 & $1/2$&$(2, 3, 6)$ 
& 1
\\ \hline
$-5$ &$\gf_4 $ & $52$  &0 & $3/5$&$(3, 4, 8)$ 
& 16/9
\\ \hline
$-6$ &$\ge_6 $ & $78$ & 0& $2/3$&$(3, 4, 8)$ 
& 8/3
\\ \hline
$-7$ &$\ge_7 $ & $133$&  $\frac{1}{2}${\bf  56}& $5/7$ &$(3, 5, 9)$ 
& 57/16
\\ \hline
$-8$ &$\ge_7 $ & $133$ & 0& $3/4$ &$(3, 5, 9)$ 
&9/2
\\ \hline
$-12$ &$\ge_8 $ & $248$ & 0&$5/6$& $(4, 5, 10)$ 
&25/3
\\ \hline
$-3, -2$ &  $\gg_2 \oplus \gsu(2) $ & $17$ & $({\bf 7} + {\bf 1},
\frac{1}{2} {\bf 2})$& $\frac25, \frac15$&$(2, 3, 6), (1, 2, 3)$ 
& 3/8
\\ \hline 
$-3, -2, -2$ &  $\gg_2 \oplus \gsu(2)$ & 17 &$({\bf 7} +{\bf 1},
\frac{1}{2} {\bf 2})$& $\frac37, \frac27, \frac17$&$(2, 3, 6), (2, 2, 4), $ 
&5/12
\\ 
& &  && & (1, 1, 2 ) 
&
\\ \hline 
$-2, -3, -2$ &  $\gsu(2) \oplus \gso(7) 
$
& 27 & $( {\bf 1},{\bf 8}, \frac{1}{2} {\bf 2})
$ & $\frac14, \frac12, \frac14$&$(1, 2, 3), (2, 4, 6),$
&1/2
\\ 
&${}\oplus \gsu(2)$& &$+( \frac{1}{2} {\bf 2},{\bf 8}, {\bf 1})$& & (1, 2, 3) 
& 
\\
\hline
\end{tabular}
\end{center}
\caption[x]{\footnotesize  Irreducible geometric components
  (non-Higgsable clusters, or ``NHC's'') consisting
of one or more intersecting curves associated with irreducible
effective divisors each with negative self-intersection.  Each cluster
including at least one curve of self-intersection $-3$ or less
gives rise to a minimal gauge algebra and matter configuration.
The numbers $\zeta_i$ are the coefficients appearing in the
Zariski decomposition of $-K$ for each cluster.}
\label{t:clusters}
\end{table}

The various clusters described in Table~\ref{t:clusters} can be
connected by $-1$ curves that intersect multiple clusters
in various ways.  The details of which clusters
can be connected by  $-1$ curves are described in \cite{clusters}.
There can also be connected sets of $-2$ curves connected to each
other and the irreducible clusters by $-1$ curves.

\begin{figure}
\begin{center}
\begin{picture}(200,70)(- 100,- 105)
\put(-190,-40){\line(1,0){80}}
\thicklines
\put(-165,-20){\line( 0,-1){30}}
\thinlines
\put(-120,-30){\makebox(0,0){$-m -1$}}
\put(-177,-25){\makebox(0,0){$-1$}}
%
\put(150,-35){\makebox(0,0){\includegraphics[width=3.5cm]{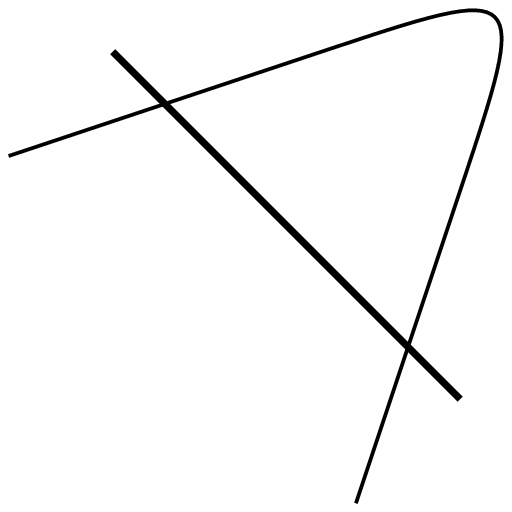}}}
\put(185,-10){\makebox(0,0){$-m-4$}}
\put(140,-55){\makebox(0,0){$-1$}}
%
\put(-15,-20){\line( -1, -1){35}}
\put(15,-20){\line(1, -1){35}}
\put(-57,-40){\makebox(0,0){$-m-1$}}
\put(55,-40){\makebox(0,0){$-n-1$}}
\put(0,-40){\makebox(0,0){$-1$}}
\thicklines
\put(-30,-27){\line(1, 0){60}}
\put(-150,-70){\vector( 0, -1){30}}
\put(150,-70){\vector( 0, -1){30}}
\put(0,-70){\vector( 0, -1){30}}
\end{picture}

\begin{picture}(200,70)(- 100,20)
\put( -165,50){\makebox(0,0){\includegraphics[width=0.3cm]{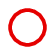}}}
\put(-190,50){\line(1,0){80}}
\thicklines
\thinlines
\put(-130,60){\makebox(0,0){$-m$}}
%
\put(150,50){\makebox(0,0){\includegraphics[width=3.5cm]{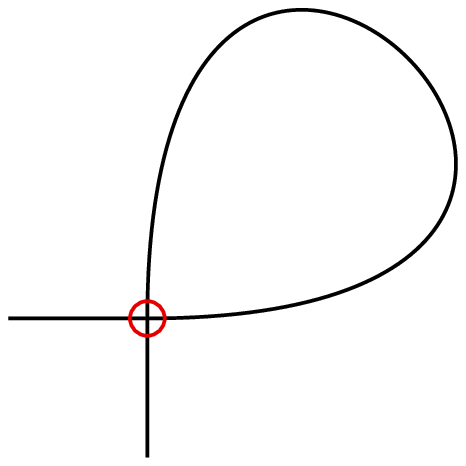}}}
\put(180,70){\makebox(0,0){$-m$}}
%
\put( 0,55){\makebox(0,0){\includegraphics[width=0.3cm]{circle.eps}}}
\put(-50,30){\line(2,1){60}}
\put(50,30){\line(-2,1){60}}
\put(34,45){\makebox(0,0){$-n$}}
\put(-34,45){\makebox(0,0){$-m$}}
\thicklines
\end{picture}
\end{center}
\caption[x]{\footnotesize Results of blowing up a point on the
  intersection structure of various configurations of irreducible
  effective curves.  Arrow indicates the map $\pi:B' \rightarrow B$
  from the blown-up space in each case.}
\label{f:blow-up}
\end{figure}

While a codimension one locus in the base where $\Delta$ vanishes to
order 12 cannot be made compatible with a resolved elliptic fibration
where the total space is a Calabi-Yau threefold, the same is not true
of a codimension two singularity locus.  If the degrees of vanishing
of $f, g, \Delta$ reach $4, 6, 12$ at a point in the base, it is still
possible to resolve the geometry to get a valid F-theory
compactification.  This requires, however, that the point in the base
$B$ is blown up, associated with an extremal transition to another
base geometry $B'$.  Blowing up a point in $B$ modifies the space of
divisors and the corresponding intersection form in a simple way
described by standard results in algebraic geometry
\cite{Griffiths-Harris, bhpv}.  
In general, blowing up a point $p$ in
the base gives a new rational curve with the topology of $\P^1$.  This
curve is a new irreducible effective divisor $E$ (the ``exceptional
divisor'') with self-intersection $-1$.  
The blow-up is described generally by a
map $\pi:B' \rightarrow B$ where $\pi:E \rightarrow p$.
Any curve $C$ with
self-intersection $C \cdot C = n$ in $B$ that passes through $p$ with
multiplicity $m$ goes to a curve $\bar{C}$ in $B'$ (the \new{proper
  transform} of $C$) with self-intersection $\bar{C} \cdot \bar{C} =
n-m^2$ and $\bar{C} \cdot E = m$.  The anticanonical divisor of $B'$
contains $E$ as a component
\begin{equation}
-K' = -\bar{K} + E \,,
\end{equation}
where $-\bar{K}$ denotes the total transform, i.e., the proper transform
of a $-K$ that does not pass through $p$.
The results of blowing up a point on one of a set of intersecting
irreducible effective curves are shown graphically in some simple
cases in Figure~\ref{f:blow-up}.  Blowing up a point on a $-m$ curve
$C$ leads to a curve $\bar{C}$ with self-intersection $-m-1$ crossed by the
exceptional $-1$ curve.  Blowing up a point at an intersection between
a  $-m$ curve and a $-n$ curve gives a chain of three curves of
self-intersection $-m-1, -1, -n-1$, with the exceptional divisor in
the middle.  These are the only cases needed in the analysis of toric
varieties that we focus on in this paper.  More generally, the
geometry can be more complicated with multiple intersections between
divisors and singularities such as nodes on the divisors.  We touch on
these issues briefly in Section \ref{sec:non-toric}

\subsection{Toric bases}

{\it Toric varieties} are
a particularly beautiful and simple class of algebraic varieties.
A toric variety is essentially an algebraic
variety carrying an action of $(\C^*)^n$, where $n$ is the complex
dimension of the variety.  Toric varieties of complex
dimension $n$ can be thought of as different compactifications of
$(\C^*)^n$.  We only use the very basic aspects of toric geometry in
this paper.  A good introduction to the subject can be found in
\cite{Fulton}, and in various papers such as \cite{Knapp-Kreuzer} that
emphasize the application to physics.

We give here a brief review of the elements of toric geometry needed
for understanding toric varieties of complex dimension two, or toric
surfaces, following \cite{Fulton}.
A toric surface is defined by a {\em fan} consisting of a set of
one-dimensional and two-dimensional cones emanating from the origin 0
of the 2D lattice $N =\Z^2$.  The 1D cones are rays generated by
integral vectors $v_i \in N$, with no two vectors parallel.  The 2D
cones are the regions between adjacent rays (which must be convex).
The origin is a zero-dimensional cone that is also included in the
fan.  The origin represents the complex torus $(\C^*)^2$, which is
contained in every toric variety, 1D cones represent divisors, and 2D
cones represent points, with the geometric picture of inclusions for
the manifold matching that of the toric diagram although dimensions
are inverted.  

The set of holomorphic functions on the set $(\C^*)^2$ associated with
the origin of $N$ is $\C[z, z^{-1}, w, w^{-1}]$.  A holomorphic
monomial is associated with every element of the dual lattice $M =$
Hom ($N,\Z$).  Choosing a specific basis $e_x, e_y$ for $N$, the
element $ae_x^*+ be_y^*$ is associated with the monomial $z^aw^b$.
For every cone $\sigma$ in the fan there is a ring of holomorphic
functions that extend to the part of the variety associated with that
cone.  This ring is described by the dual cone $\sigma^*=\{u\in M: 
\langle u,
v \rangle\geq 0 \; \forall v\in \sigma\}$.  This gives a set of local coordinate
patches that can be glued together giving a global algebraic
description of the variety.  The surface is compact when the fan
covers the entire plane --- {\it i.e.}, when the 2D cones are taken
between all pairs of vectors in a cyclic ordering.  The surface is
smooth if the vectors bounding each 2D cone form a basis for the
lattice.

As a simple example, the fan for $\P^2$ is generated by the vectors
\begin{equation}
\P^2: \; v_1 = (1, 0), v_2 = (0, 1), v_3 = (-1, -1) \,.
\label{eq:toric-p2}
\end{equation}
The holomorphic functions on the torus $(\C^*)^2 \subset \P^2$ are
spanned by monomials $z^nw^m$, with $(n, m) \in M =\Z^2$.  For the cone
spanned by $v_1, v_2$ the dual cone is spanned by $(0, 1), (1, 0)$ so
in that coordinate patch local holomorphic functions are $z^n w^m, n,
m \geq 0$, generated by $z, w$.  In the patch associated with the cone
spanned by $v_1, v_3$ the holomorphic functions are generated by
$w^{-1}, zw^{-1}$, and in the cone spanned by $v_1, v_2$ the
holomorphic functions are generated by $z^{-1}, wz^{-1}$.  These are
the usual local coordinates on $\P^2$ in homogeneous coordinates
$(s:t:u)$ where $z = s/u, w = t/u$, with the usual rules for patching
together the different local coordinate charts.
As another class of examples, the
fan for the Hirzebruch surface $\F_m$ is generated by the vectors 
\begin{equation}
\F_m: \; v_1 = (1, 0), v_2 = (0, 1), v_3 = (-1, -m), v_4 = (0, -1) \,.
\label{eq:toric-fm}
\end{equation}
The Hirzebruch surfaces are $\P^1$ bundles over $\P^1$; this can be
seen in the toric description, from the projection along the vertical
axis; the vectors $v_2, v_4$ provide the fan of the fiber $\P^1$, and the
projection of the remaining vectors provide the fan of the base $\P^1$.

The irreducible effective divisors and intersection structure of a
toric variety can be read off directly from the toric diagram.  The
general algorithm for deriving the intersection ring from the toric
data is particularly simple in the case of surfaces.  On a surface,
each vector $v_i$ generating a 1D cone represents an irreducible
effective divisor $D_i$.  There are two linear relations on these
divisors, given by
\begin{equation}
\sum_{i} \langle
m, v_i \rangle D_i = 0
\label{eq:linear-relations}
\end{equation}
for  $m \in M$.  Each pair of adjacent divisors has an
inner product $D_i \cdot D_j = 1$.  For a smooth surface, each vector
$v_i$ satisfies $v_{i -1} + v_{ i+1} = n_i v_i$ for some $n_i$; the
self intersection of the corresponding divisor is $D_i \cdot D_i =
-n$.  The negative of the canonical class $K$ of  any toric variety is given by
the sum of divisors associated with all $v_i$
\begin{equation}
-K = \sum_{i}D_i \,.
\label{eq:toric-k}
\end{equation}
For example, for the Hirzebruch surfaces the linear relations give $F
\equiv D_1 \sim D_3$ and $S_\infty \equiv D_4 \sim D_2 -mD_3$ as the generators
of the cone of effective divisors (Mori cone), where $F$ is a fiber
and $S_\infty$ is the section of the elliptic fibration, with $F \cdot F = 0,
F \cdot S_\infty= 1, S_\infty\cdot S_\infty= -m$.  The irreducible effective divisor
$S_0= S_\infty+ mF = D_2$ is a generic section with $S_0 \cdot
S_0= m$.  The canonical class for the Hirzebruch surfaces is $
-K = \sum_{i}^{}D_i = 2S_\infty+ (2 + m)F$.

\begin{figure}
\begin{center}
\begin{picture}(200,160)(- 100,- 80)
\put(-100,10){\makebox(0,0){\includegraphics[width=5cm]{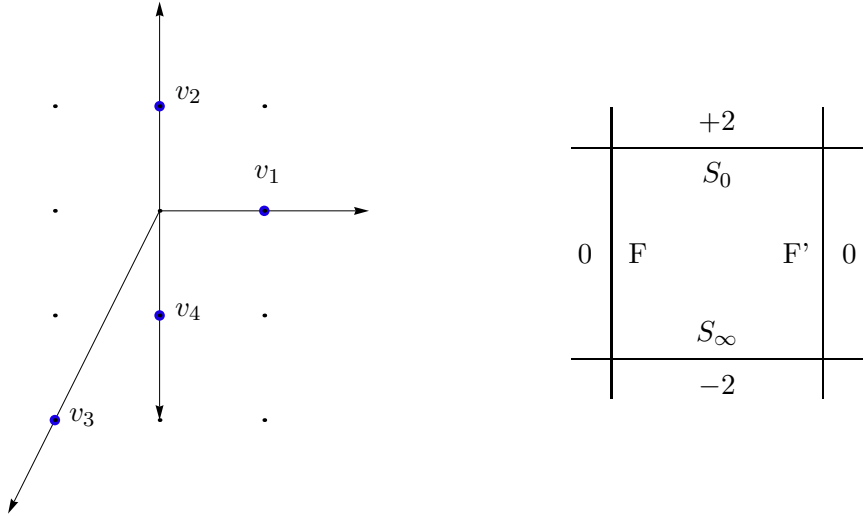}}}
\put(-70,30){\makebox(0,0){$v_1$}}
\put(-100, 60){\makebox(0,0){$v_2$}}
\put(-140, -62){\makebox(0,0){$v_3$}}
\put(-100, -22){\makebox(0,0){$v_4$}}
\put(45, 40){\line(1, 0){110}}
\put(45, -40){\line(1, 0){110}}
\put(60, 55){\line( 0, -1){110}}
\put(140, 55){\line( 0, -1){110}}
\put(100, 50){\makebox(0,0){$+2$}}
\put(100, -50){\makebox(0,0){$-2$}}
\put(50,0){\makebox(0,0){$0$}}
\put(150,0){\makebox(0,0){$0$}}
\put(100, 30){\makebox(0,0){$S_0$}}
\put(100, -30){\makebox(0,0){$S_\infty$}}
\put(70,0){\makebox(0,0){F}}
\put(130,0){\makebox(0,0){F'}}
\end{picture}
\end{center}
\caption[x]{\footnotesize Toric diagram and corresponding loop of
  connected curves representing irreducible effective divisors for
  Hirzebruch surface $\F_2$.}
\label{f:loop-examples}
\end{figure}

The set of irreducible effective curves associated with the toric
divisors can be described graphically as a loop of connected curves,
as shown in the right side of Figure~\ref{f:loop-examples} (see also
Figure~\ref{f:example-12}).
For the toric surface to be a good base for an F-theory
compactification, the sequence of intersection numbers must correspond
to an allowed sequence of blocks from Table~\ref{t:clusters} 
connected
by curves of self-intersection $-1$ or less.
In \cite{clusters} a complete analysis is given of which blocks can be
connected by a curve of self-intersection $-1$.  The results of this
analysis relevant for toric surfaces are summarized in
Table~\ref{t:connections}.  For example, a $-4$ curve can be connected
to another $-4$ curve by a $-1$ curve, but not to a $-5$ curve as the
degrees of vanishing of $f, g, \Delta$ become too great at the
intersection of the $-1$ and $-5$ curves, so that point must be blown
up to get an acceptable F-theory base.  These connectivity rules
provide strong constraints on which toric surfaces can be used as
F-theory bases.

\begin{table}
\begin{center}
\begin{tabular}{| c | c  | 
}
\hline
Cluster & Possible subsequent clusters
\\
\hline
(-12) & (-2, -2, -3)\\
(-8) & (-2, -3, -2) or below\\
(-7) & (-2, -3, -2) or below\\
(-6) & (-3) or below\\
(-5) & (-3, -2, -2) or below\\
(-4) & (-4) or below\\
(-3, -2, -2) & any cluster\\
(-3, -2) & (-8) or below\\
(-3) & (-6) or below\\
(-2, -3, -2) & (-8) or below\\
(-2, -3) & (-5) or below\\
(-2, -2, -3) & (-5) or below\\
(-2, -2, \ldots, -2) & any cluster\\
\hline
\end{tabular}
\end{center}
\caption[x]{\footnotesize Allowed connections between non-Higgsable
  clusters by $-1$ curves in a toric surface that can be used as an
  F-theory base.  For each cluster, the table indicates the clusters that
  can follow the first cluster after a $-1$ curve, where ``or below''
  refers to the order of clusters in this table.  Note that the
  clusters $(-3, -2, -2)$ and $(3, 2)$
are ordered; for example, a $-12$ can be
  connected by a $-1$ curve to the final $-2$ of the cluster $(-3, -2,
  -2)$ but not
  to the $-3$ curve.  For clarity these clusters are listed in both
  directions.}
\label{t:connections}
\end{table}

\begin{figure}
\begin{center}
\begin{picture}(200,320)(- 100,- 160)
\put(-120, 120){\makebox(0,0){\includegraphics[width=4cm]{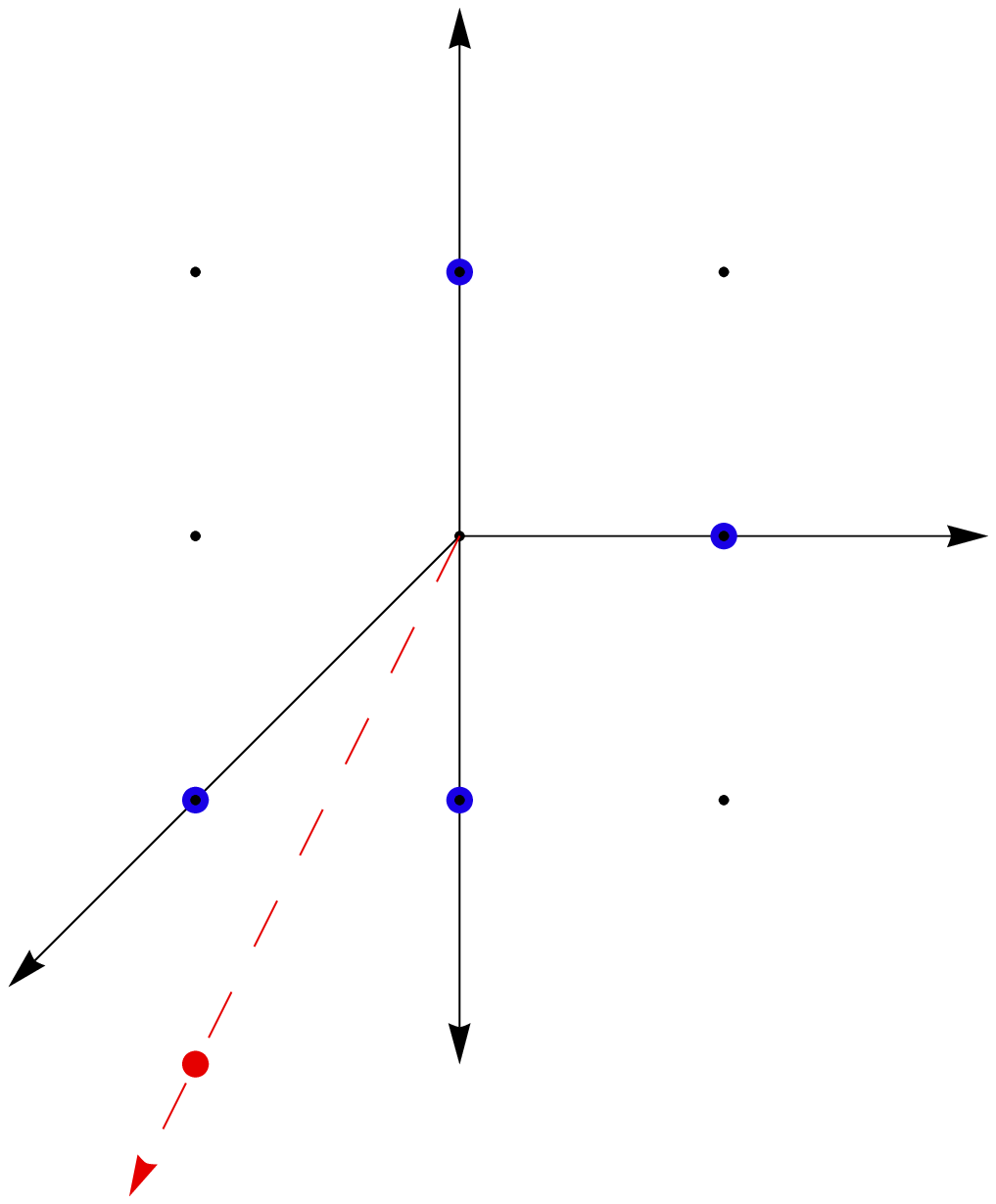}}}
\put(-120,  -100){\makebox(0,0){\includegraphics[width=4cm]{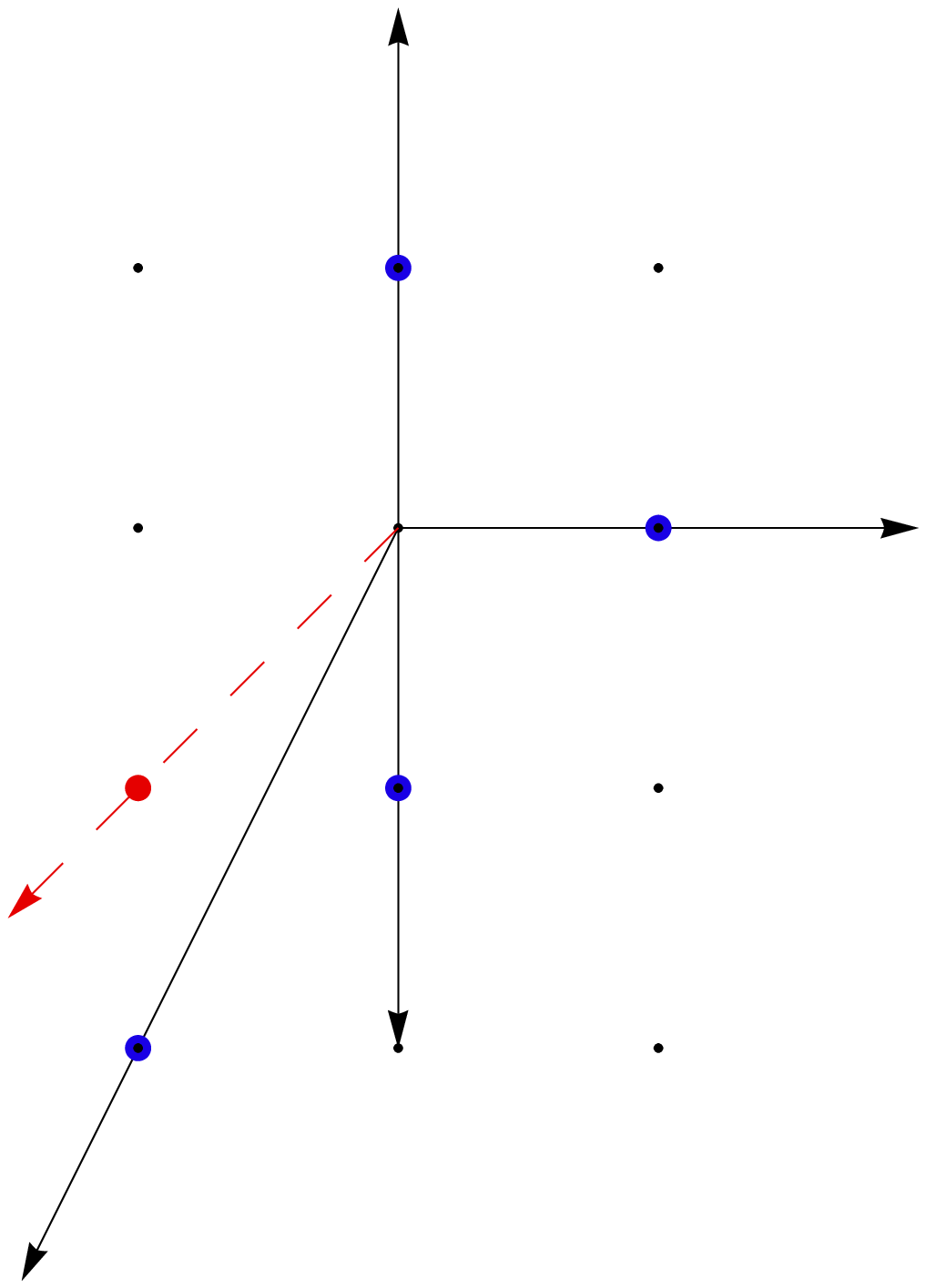}}}
\put(70,  15){\makebox(0,0){\includegraphics[width=4cm]{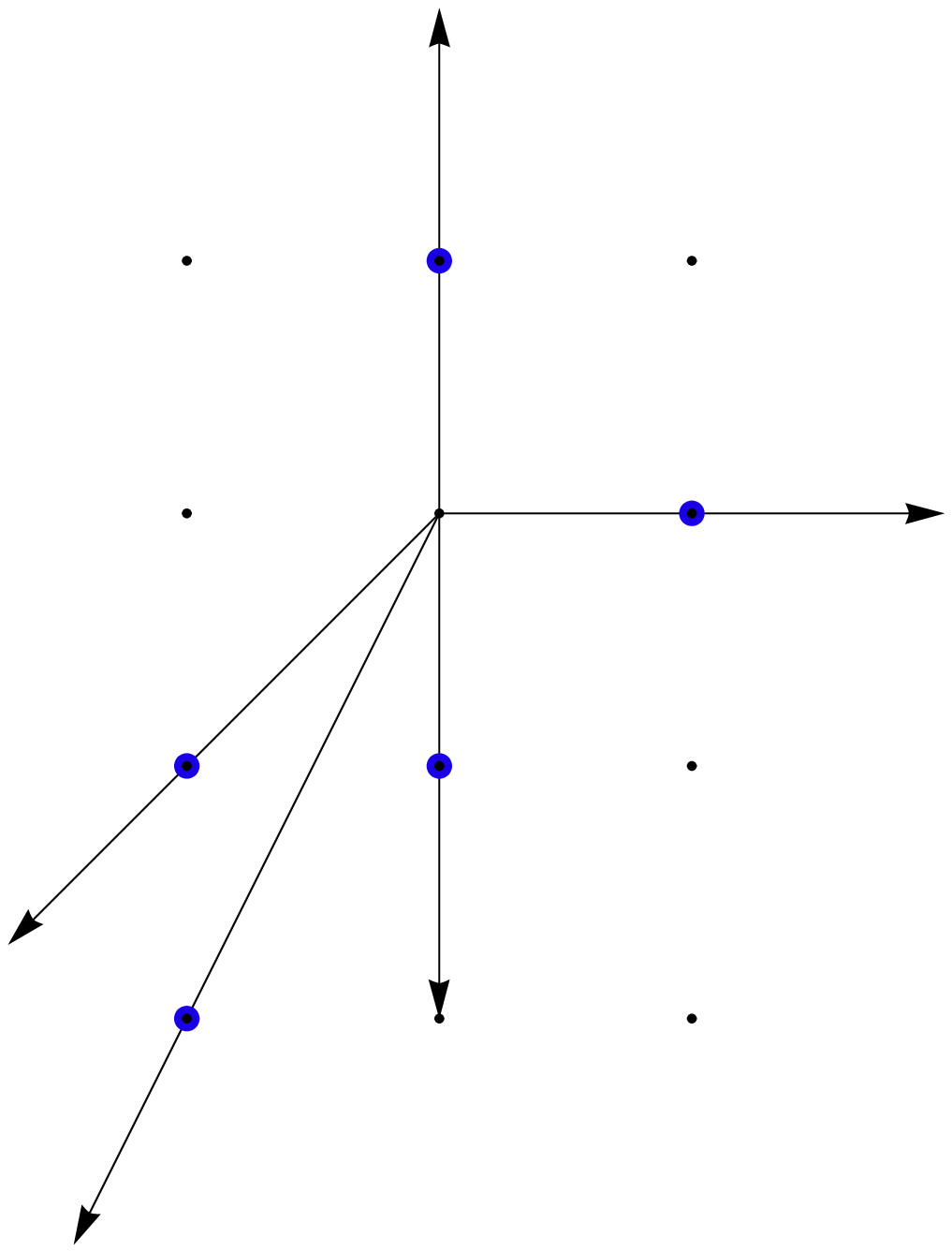}}}
\put(-50,110){\makebox(0,0){\includegraphics[width=0.3cm]{circle.eps}}}
\put(-50,-100){\makebox(0,0){\includegraphics[width=0.3cm]{circle.eps}}}
\put(-60,150){\line(1,0){60}}
\put(-60,110){\line(1,0){60}}
\put(-60,-100){\line(1,0){60}}
\put(-60,-140){\line(1,0){60}}
\put(-50,160){\line(0,-1){60}}
\put(-50,-90){\line(0,-1){60}}
\put(-10,160){\line(0,-1){60}}
\put(-10,-90){\line(0,-1){60}}
\put(145,50){\line(1, 0){55}}
\put(145,10){\line(1, 0){55}}
\put(140,20){\line(1,2){20}}
\put(140,40){\line(1,-2){20}}
\put(190,60){\line(0,-1){60}}
\put(-30,158){\makebox(0,0){\small $+1$}}
\put(-30,102){\makebox(0,0){\small $-1$}}
\put(-30,-92){\makebox(0,0){\small $+2$}}
\put(-30,-148){\makebox(0,0){\small $-2$}}
\put(-58,130){\makebox(0,0){\small $0$}}
\put(-58,-120){\makebox(0,0){\small $0$}}
\put(-2,130){\makebox(0,0){\small $0$}}
\put(-2,-120){\makebox(0,0){\small $0$}}
\put(198, 30){\makebox(0,0){\small $0$}}
\put(172, 58){\makebox(0,0){\small $+1$}}
\put(172, 2){\makebox(0,0){\small $-2$}}
\put(142, 18){\makebox(0,0){\small $-1$}}
\put(142, 42){\makebox(0,0){\small $-1$}}
\put(-120,50){\makebox(0,0){$\F_1$}}
\put(-120,-160){\makebox(0,0){$\F_2$}}
\put(70,-55){\makebox(0,0){$X_1$}}
\thicklines
\put(-10, -30){\vector( -2,-1){40}}
\put(-10,70){\vector( -2,1){40}}
\end{picture}
\end{center}
\caption[x]{\footnotesize Blow-up and blow-down transitions connect
the surfaces that can be used as F-theory bases.  Examples of
blow-ups connecting several toric bases are shown.
Dashed
(red) vectors and circled vertices represent points blown up on $\F_1$ and $\F_2$ that give a common toric base with $\ho (B) = 3$.}
\label{f:toric-connections}
\end{figure}

A point represented by a 2D cone $\sigma_{ij}$ bounded by 1D cones generated by
$v_i, v_j$ can be blown up by adding a new
vector $v = v_i + v_j$ to the fan, and dividing the 2D cones
accordingly.  Similarly, a curve represented by a vector $v$ which is
equal to the sum of the adjacent vectors is always a $-1$ curve and
can be blown down.  For
example, $\P^2$ as described in \eq{eq:toric-p2}
can be blown up at the point in the cone $\sigma_{13}$
bounded by $v_1, v_3$ giving
the point $v_4 = (0, -1)$, reproducing the toric description
\eq{eq:toric-fm} of $\F_1$.  Since $\P^2$ admits a symmetry which maps
any point to any other, this is the only blow-up of $\P^2$.
For each $\F_m, m > 0$, there are two possible blow-ups.  We can blow up the
cone $\sigma_{23}$, adding the vector $v = (-1, -m + 1)$ or we can
blow up the cone $\sigma_{34}$, adding the vector $v = (-1, -m -1)$.
(Note that blowing up the cones $\sigma_{12}, \sigma_{41}$ give
equivalent results after a redefinition of lattice basis).  The
blow-up of $\sigma_{23}$ corresponds to the blow-up of $\F_m$ at a
generic point, while the blow-up of $\sigma_{34}$ corresponds to
blowing up $\F_m$ at a point on the divisor $D$ associated with $v_4$,
which satisfies $D \cdot D = -m$.  The results of blowing up an
$\F_m$ in either way describe a class of surfaces 
$X_m$, with toric generators
\begin{equation}
X_m: \; v_1 = (1, 0), v_2 = (0, 1), v_3 = (-1, -m), 
v_4 = (-1, -m-1), v_5 = (0, -1) \,.
\label{eq:toric-xm}
\end{equation}
Thus, blow-ups connect $\F_m$ to $X_m$ and $X_{m-1}$.
These blow-ups are shown in both toric and graphical curve form in
Figure~\ref{f:toric-connections}.  

\section{Enumeration of bases}
\label{sec:toric}

The mathematical classification of complex surfaces is well
understood \cite{bhpv,Reid-chapters}.  In the \new{minimal model} approach,
a surface
is systematically reduced by blowing down $-1$ curves until a minimal
surface without $-1$ curves is reached.  Using this approach, all
smooth F-theory bases giving 6D ${\cal N} = 1$ supergravity theories
can be blown down to $\P^2$ or $\F_m$ by successfully blowing down
irreducible effective $-1$ curves \cite{Grassi91,
  Morrison-Vafa-II}.\footnote{The Enriques surface is also a possible
  F-theory base; this branch gives rise to a theory with no gauge
  structure or matter content.  (Note that the Enriques surface cannot
be further blown up without destroying the Calabi--Yau property of
the total space of the elliptic fibration.)  It may also be possible to make sense
  of F-theory on base spaces with orbifold singularities
  \cite{Morrison-Vafa-II}; we do not consider such bases
  here.}
Note that after blowing down an F-theory model, 
the Weierstrass model on the blown down surface has $(f,g,\Delta)$
of multiplicities at least $(4,6,12)$ at the blown down points, which
is why we normally study these models in blown-up form.

This minimal model result implies
that all smooth toric F-theory bases can be reached by starting
with $\P^2$ or $\F_m$ and successively blowing up points at the
intersection between adjacent divisors to generate bases with
increasing values of $T$.  After a sequence of such blow-ups, only
those bases described by a sequence of intersection numbers that is
built from clusters listed in Table~\ref{t:clusters} connected by
curves of self-intersection $-1$ or greater with the connections obeying the
rules in Table~\ref{t:connections} are possible.  We have
systematically enumerated all possible toric bases for 6D F-theory
models using this approach.  The computational details of this
calculation are described in the Appendix.  The result is that we find
61,539 distinct smooth toric bases possible for 6D F-theory
compactifications.  Some of these toric bases contain $-9, -10,$ or
$-11$ curves with additional singularities that must be blown up where
they cross the discriminant locus.  While these bases are not
technically toric after this blow-up, they have a closely related
structure and we include them in our analysis; we discuss this point
further below.  Taking a strict definition of toric bases, by not
including the bases with these types of curves, reduces the total
number of bases to 34,868, with the largest value of $T$ being 129.
The values of $T$ for the allowed bases range from $\P^2$ and $\F_m$
with $T = 0, 1$ respectively, to a base with $T = 193$, where in
the last case the toric base has two $-11$ curves that are blown up,
as discussed further in Section~\ref{sec:large}.

\subsection{Distribution of bases, non-Higgsable gauge groups, and
  matter content}

\begin{figure}
\begin{center}
\includegraphics[width=10cm]{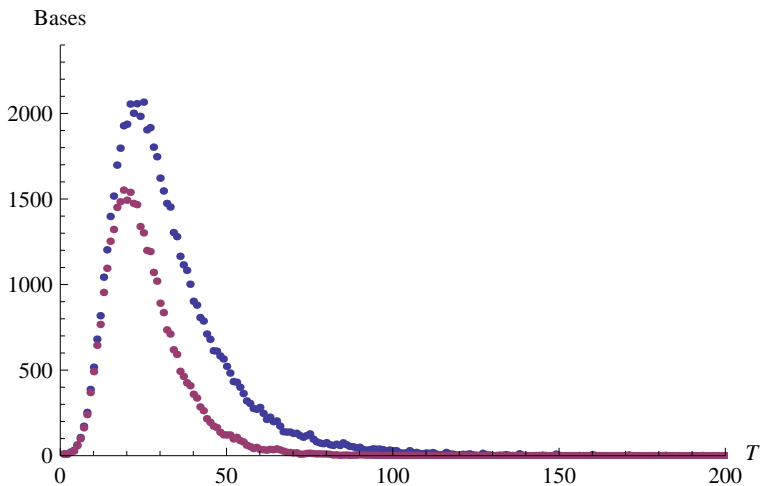}
\end{center}
\caption[x]{\footnotesize The number of distinct toric bases for
  F-theory compactifications for different numbers $T$ of tensor
  multiplets.  There are 61,539 toric bases including those with $-9,
  -10, -11$ curves that must be blown up (upper blue data), and 34,868
  truly toric bases not including such curves (lower purple data).
  The largest number of tensor multiplets is $T = 193$.}
\label{f:numbers}
\end{figure}

The number of bases identified for each possible number of tensor
multiplets $T$ is graphed in Figure~\ref{f:numbers}.  The numbers of
bases for some representative values of $T$ are listed in
Table~\ref{t:numbers}.  For $T = 0,1, 2$ the bases are $\P^2$, the
Hirzebruch surfaces $\F_m$ and the surfaces $X_m$ described above.
(Note that the Hirzebruch surfaces $\F_m$ contain singular points that
must be blown up for a good F-theory base at $m = 9, 10, 11$, so that
there are only 10 good F-theory bases at $T = 1$.)  The largest number
of distinct bases appears at $T = 25$, where there are 2066 distinct
toric bases.  The range of bases rapidly drops between $T =$ 30 and
60, and $T = 141$ is the smallest value of $T$ with no allowed toric
bases.  The bases become more sporadic up to the maximum of $T = 193$.
Some details of the bases with large $T$ are discussed in the
following subsection.

\begin{table}
\begin{center}
\begin{tabular}{| c | c | c  | c | c  | c | c  ||
c  || c  || c  || c || c || c  || c  || c | 
}
\hline
0 &1 & 2 & 3 & 4 & 5 & 6 & 10 & 20 &25 &30 &50 &100 & 150 & 193\\
\hline
1 &10 &  10 &  21 & 31 & 63 & 106 & 517 &  1937 & 2066 &  1622 & 522 &29
&  2 & 1\\
\hline
\end{tabular}
\end{center}
\caption[x]{\footnotesize  Numbers of distinct toric bases for some
  sample values of $T =\ho-1$}
\label{t:numbers}
\end{table}

\begin{figure}
\begin{center}
\begin{picture}(200,180)(- 100,- 90)
\put(-110,0){\makebox(0,0){\includegraphics[width=7cm]{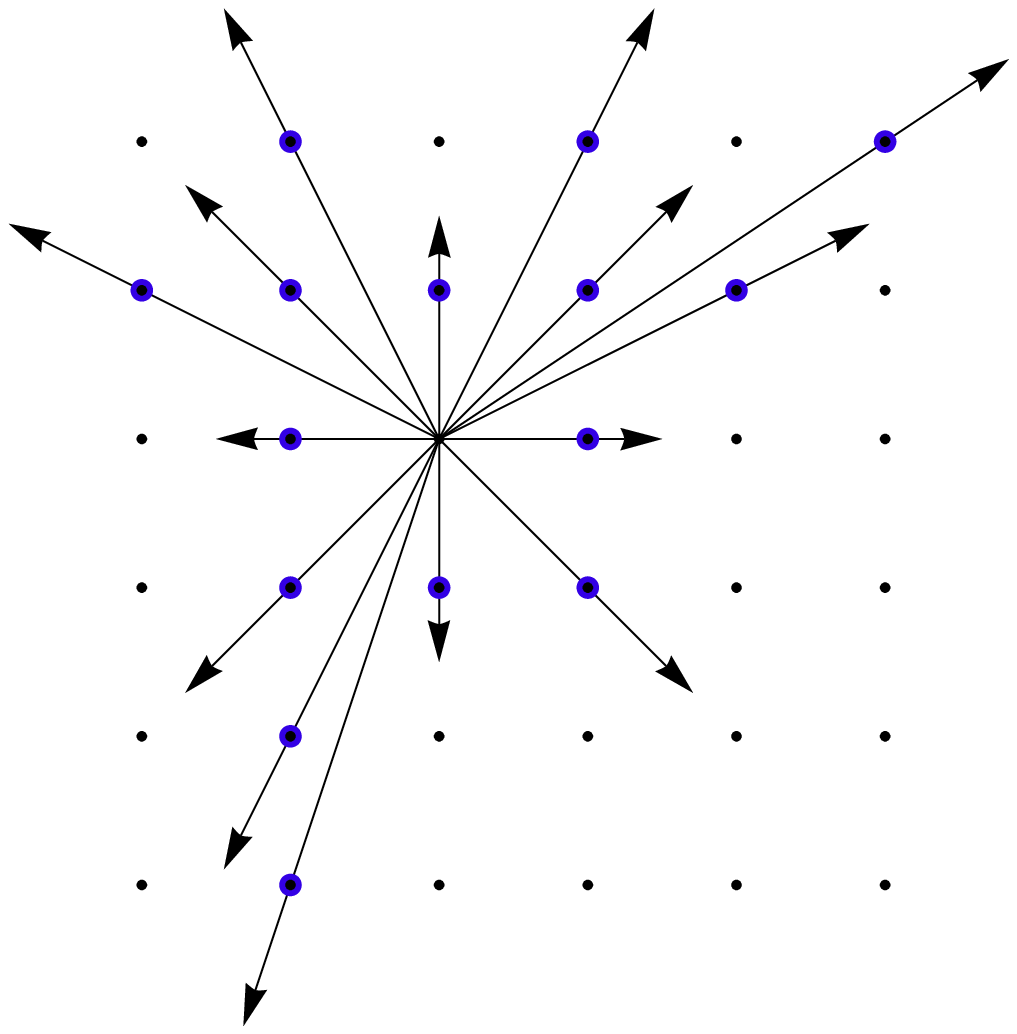}}}
\put(110,0){\makebox(0,0){\includegraphics[width=7cm]{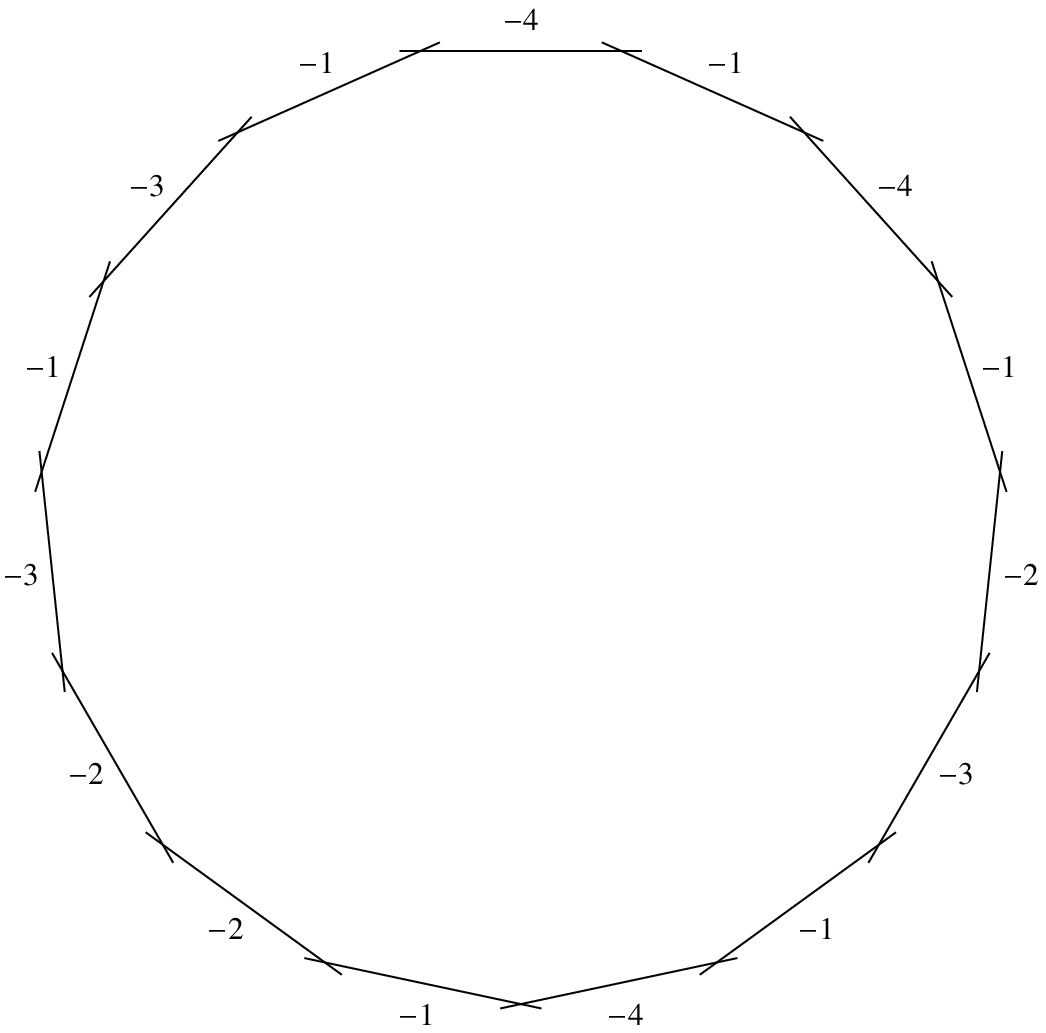}}}
\end{picture}
\end{center}
\caption[x]{\footnotesize A typical toric F-theory base with $\ho =
  13$, corresponding to a 6D supergravity theory with $T = 12$.  The
  sequence of self-intersections of the curves around the loop
is given by (-4, -1,
  -4, -1, -2, -3, -1,-4, -1, -2, -2, -3, -1, -3, -1).  This base
gives a 6D supergravity theory with non-Higgsable gauge algebra 
$\gso(8) \oplus\gso(8) \oplus\gso(8) \oplus (\gg_2 \oplus
\gsu(2))\oplus (\gg_2 \oplus
\gsu(2)) \oplus \gsu(3)$.}
\label{f:example-12}
\end{figure}

A fairly typical\footnote{Note, most bases at $T = 12$ have at least
  one divisor of self-intersection $-6$ or less, so this is not a
  completely typical base.} base at $T = 12$ is depicted in
Figure~\ref{f:example-12}, both in toric fan and
graphical curve form.
Any toric base is characterized simply by the sequence of self-intersection
numbers $C_i^2$ of the irreducible effective curves that form a closed
loop.  From the clusters of
curves with self-intersections below $-1$ we can read
off the non-Higgsable gauge algebra content of the supergravity model
associated with any such base.  For example, for the base depicted in
Figure~\ref{f:example-12} the gauge algebra is
\begin{equation}
\gso(8) \oplus\gso(8) \oplus\gso(8) \oplus (\gg_2 \oplus
\gsu(2))\oplus (\gg_2 \oplus
\gsu(2)) \oplus \gsu(3)\,.
\end{equation}

One subtlety, noted above, is that many of the toric bases
reached after blowing up a sequence of points on a surface $\F_m$
contain curves $C$ of self-intersection $-m =-9, -10,$ and $-11$.  In
these cases, as mentioned in \cite{clusters}, points on these curves
must be blown up for a good F-theory model.  Specifically,
the degree of vanishing of
the discriminant locus $\Delta = -12K$ on $C$ is given by $d =[\Delta]
=\lceil 12 (m-2)/m\rceil = 10$ from $-K \cdot C = -m + 2$.  Similarly,
the degrees of vanishing of $f, g$ are $[f] = 4,[g] = 5$.  
(This can easily be read off from the Zariski decomposition given above.)
The residual part of the discriminant locus $Y = -12K-dC$ has a further
intersection with $C$ of multiplicity $24 -2m$.  Writing the equation
of the curve $C$ as $z = 0$, we see locally that $f = z^4 \tilde{f}, g
= z^5 \tilde{g}$, so $\Delta = z^{10} (4 z^2 \tilde{f}^3 + 27
\tilde{g}^2)$.  It follows that the residual discriminant locus has a
factor of $(4 z^2 \tilde{f}^3 + 27 \tilde{g}^2)$ and is locally
tangent to $C$.  At these intersection points the degree of $\Delta$
is increased to 12, so the points must be blown up.  After blowing up
these points we have curves of self-intersection $-12$ that carry an
$\ge_8$ gauge algebra without matter.  Physically, this resolution arises
because there is no appropriate representation for matter charged
under the $\ge_8$ where it intersects the residual curve.  While the
bases that result from this process are not actually toric, due to
the final blow-ups on the $-9, -10$ or $-11$ curves, the structure of
these bases is closely related to that of the toric bases and we include
them in our analysis, associated with the value of $T$ that results
after the necessary blow-ups of the $-9, -10$ or $-11$ curves. 
In the remainder of this paper we use the term ``toric base'' loosely
to include both the genuine toric bases and those where $-9, -10,$ and
$-11$ curves on a toric base have been blown up.

In other work, another approach has been taken to understanding
elliptically fibered Calabi-Yau threefolds for 6D theories from toric
geometry.  Many Calabi-Yau threefolds can be realized as hypersurfaces
in toric varieties of one higher dimension, with a simple description
in terms of reflexive polyhedra using the Batyrev construction
\cite{Batyrev}.  In \cite{Candelas-pr}, threefolds for 6D F-theory
compactifications were analyzed from this approach and a method was
given for computing the gauge group and number of tensors from the
toric data.  Kreuzer and Skarke \cite{Kreuzer-Skarke} have
systematically identified the roughly 500 million reflexive polytopes
in four dimensions.  In principle, one could identify all toric elliptic
fibrations for F-theory from among this list.  Recently, Braun
\cite{Braun} has identified those reflexive polytopes that correspond
to elliptic fibrations over $\P^2$.  By focusing on the structure of
the base, the analysis we have used here simplifies the problem of
identifying the distinct possible F-theory bases.  For each base,
there will be many elliptic fibrations in which the gauge group is
enhanced, with a wide range of matter structures coming from different
types of codimension two singularities.  Braun identified 102,581
elliptic fibrations over $\P^2$.  Some of the range of possible matter
representations that can appear in 6D F-theory models over $\P^2$ are
identified from anomaly constraints in the low-energy supergravity
theory in \cite{0}, and analyzed from the point of view of codimension
two singularities in F-theory in \cite{Morrison-Taylor}.
Understanding and classifying more completely the range of structure
possible over a given base such as $\P^2$ is an interesting problem
for future work.  For each of the 61,539 toric bases we have found,
there will be a similar broad class of constructions.  As $T$
increases, however, the number of available moduli to tune additional
gauge groups decreases, as we discuss in Section
\ref{sec:Weierstrass}.  Thus, we expect that while there are many
models at larger values of $T$, the range of possibilities for
enhanced gauge groups and matter structure will decrease as $T$
increases.
An interesting project for future work would be to relate the set of
bases found here by direct toric analysis of the base surface to those
bases that appear in the elliptic fibrations from the Kreuzer and
Skarke database.

\subsection{Bases with large $T$}
\label{sec:large}

\begin{figure}
\begin{center}
\includegraphics[width=12cm]{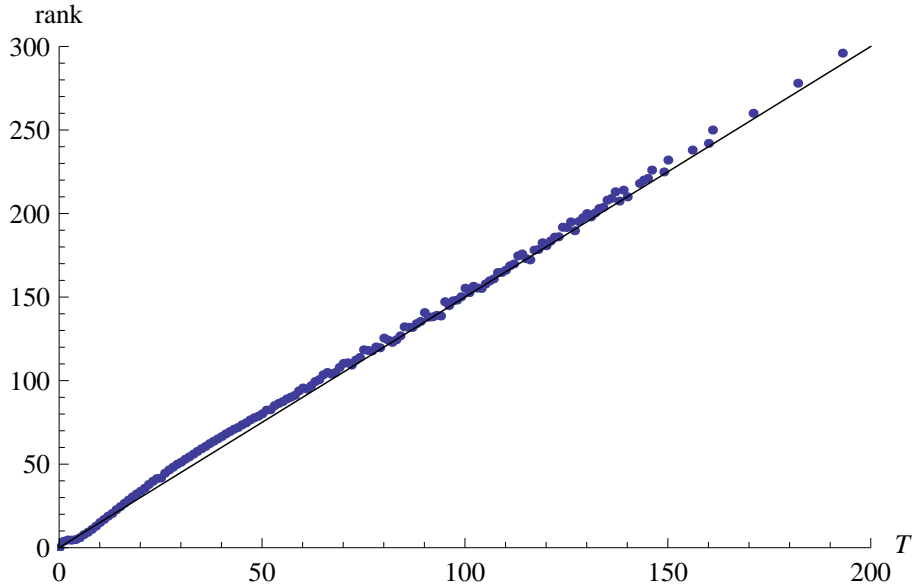}
\end{center}
\caption[x]{\footnotesize Average rank of gauge group as a function of
$T$, compared to the linear estimate  $3T/2$ (solid line
in black).}
\label{f:rank}
\end{figure}

\begin{figure}
\begin{center}
\includegraphics[width=12cm]{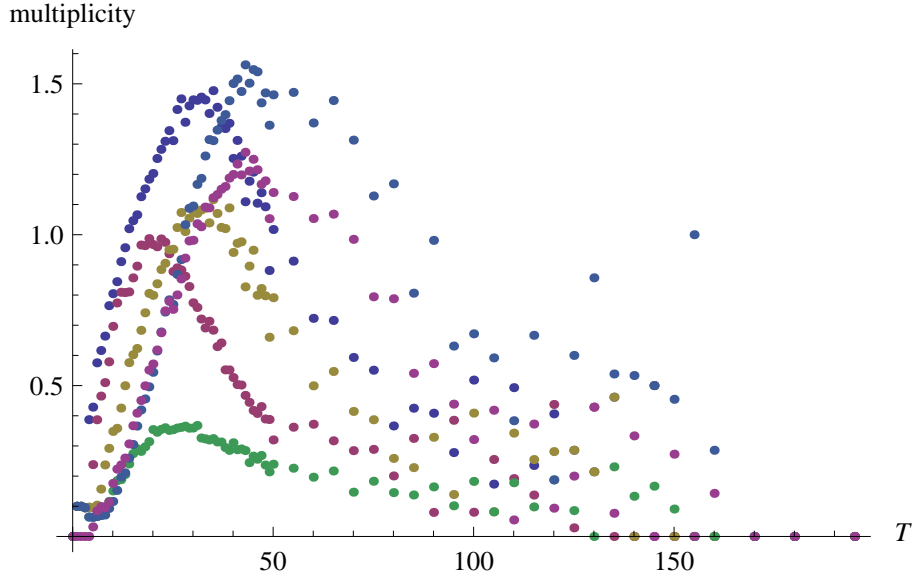}
\end{center}
\caption[x]{\footnotesize Average number of gauge algebra summands as
  a function of $T$ for summands (ordered from top at $T = 20$)
  $\gsu(3),\gso(8),\ge6,\gso(7)\oplus\gsu(2)\oplus\gsu(2),\ge_7$ (no
  matter), $\ge_7$ (non-Higgsable charged matter).  Data is binned
  into groups of 5 above $T = 50$ due to small statistics.
The point of this plot is that the average number of all of these factors is
small, peaking at most at 1.5, then dropping with increasing $T$.}
\label{f:algebras-1}
\end{figure}

\begin{figure}
\begin{center}
\begin{picture}(200,220)(- 100,- 110)
\put(0,0){\makebox(0,0){\includegraphics[width=12cm]{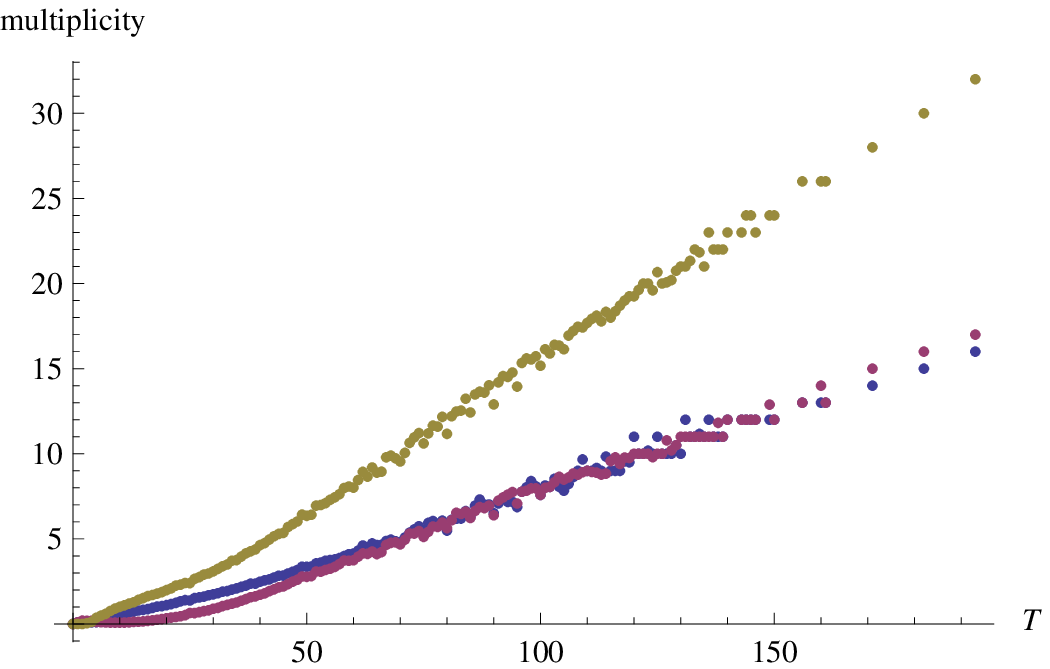}}}
\put(100,80){\makebox(0,0){$\gg_2 \oplus\gsu (2)$}}
\put(140,15){\makebox(0,0){$\ge_8$}}
\put(140,-25){\makebox(0,0){$\gf_4$}}
\end{picture}
\end{center}
\caption[x]{\footnotesize Average number of  gauge algebra summands
as a function of $T$ for summands 
$\gg_2 \oplus\gsu (2)$,
$\gf_4,\ge_8$.
For these summands, the contribution grows linearly in $T$.}
\label{f:algebras-2}
\end{figure}

As the complexity of the bases increases with $T$, the minimal
(non-Higgsable) gauge algebra associated with the bases becomes
larger.  In Figure~\ref{f:rank} we show the average rank of the gauge
group as a function of $T$.  The rank grows roughly linearly with $T$,
with a slope close to $3/2$.
To understand this it is helpful to consider the more detailed
structure of toric bases at large $T$.  In Figures~\ref{f:algebras-1}
and~\ref{f:algebras-2} we
graph the average number of each of the non-Higgsable gauge algebra
summands as a function of $T$.  As $T$ increases, the gauge algebra is
dominated by $\ge_8, \gf_4,$ and $\gg_2 \oplus \gsu(2)$ summands.  The
reason that these algebra components dominate can be understood from
several aspects of the analysis given in \cite{clusters}.  In that
paper we identified a bound on $T$ for any theory with a given
non-Higgsable gauge group.  For theories where the gauge group can be
fully Higgsed, $T$ is less than or equal to 9.  Each gauge algebra
summand contributes to the upper bound a quantity given by the last
column in Table~\ref{t:clusters}.  Thus, for example, a theory with
gauge group $SU(3)^k$ can have a value of $T$ that is at most $T \leq
9 + k/3$.
As discussed in \cite{clusters},
for most of the combinations of gauge algebra summands that appear in
linear chains that can be blown down without ``breaking'' (reducing to
sequences containing multiple curves with self-intersection $\leq -3$ and no $-1$ curves), the bound on $T$
is relatively low.  By blowing up a sequence of $-2$ curves, however
it is possible to generate a periodic sequence of the
form
\begin{equation}
\ldots, -12, -1, -2,  -2, -3,-1, -5, -1, -3, -2, -2, -1, -12, \ldots
\label{eq:pattern-12}
\end{equation}
This sequence gives the largest possible contribution to $\Delta T$
per unit length of any allowed periodic sequence.  It is natural
therefore to expect that this sequence will play an important  role in toric bases at
large $T$.
Indeed,
the toric bases with large $T$ are described by sequences of curves
dominated by this periodic pattern.  This corresponds in
Figure~\ref{f:algebras-2}
to the fact that
the number of $\ge_8$ and $\gf_4$ summands grows as $T/12$ while the number
of $\gg_2 \oplus \gsu(2)$ summands grows as $T/6$ at large $T$.
Similarly, the rank contribution from each iteration of this pattern
is 18, which with  $T$ increasing by 12 reproduces the observed
3/2 slope seen in Figure~\ref{f:rank}.

The largest toric base has 194 curves corresponding to the rays in the
toric fan.  The self-intersection numbers of these curves are
essentially 16 repeated copies of (\ref{eq:pattern-12}) with a $-12$
on each end, and a curve of self-intersection 0 closing the loop.  The
actual toric base has two $-11$ curves in the next-to-last places for
$-12$ curves.  As discussed at the end of the previous section, these
must be blown up, so that the final associated F-theory base is not strictly
toric and has $-1$ curves intersecting the second and penultimate
$-12$'s in the chain.  The resulting geometry of the F-theory base
gives the 6D
supergravity theory with the largest gauge group rank known, previously
identified in \cite{Candelas-pr, Aspinwall-Morrison-instantons}.  This
F-theory
base has $\ho = 194$ and
gives a theory with
$T = 193,$ the largest number of tensor multiplets known for any
consistent 6D supergravity theory.  
One way to reach this
geometry is to start with $\F_{12}$, blow up the intersection point
between the $+ 12$ divisor ($S_0$) and a particular fiber 24
times, and then blow up points on that fiber, leaving another fiber as
the $0$-curve that closes the loop.  Another possibility is to start
with $\F_0$, and then blow up the intersections connecting 
the $0$
curve ${S_\infty}$ with two different fibers 6 times
each, and then blow up points on the two fibers, leaving $S_0$
as the $0$-curve that closes the loop.  The first of
these possibilities  can be
described in terms of the systematic algorithm used for constructing
all toric bases in the Appendix, and
was
used in \cite{Aspinwall-Morrison-instantons} to construct the base
geometry with $T = 193$.

As the number of tensors approaches the maximum of $T = 193$ for toric
models, bases appear only sporadically.  The next-highest value of $T$
where there is a toric base is $T = 182$, where there are two
possibilities.  These bases have essentially the same structure as the
$T = 193$ base, but only 15 copies of the cycle \eq{eq:pattern-12},
and minor variations on the detailed structure.  In one case, the
$-12$'s on the end of the sequence are connected by a pair of $-1$
curves instead of a single curve of self-intersection $0$.  In the
other case, both the next-to-last and the final $-12$ on one end of
the chain are replaced by $-11$'s that are blown up at one point.  At
$T = 171$ there are four variations on the pattern with 14 copies of
the cycle \eq{eq:pattern-12}.  At $T = 161$ and below there are bases
at most values of $T$, which begin to break the pattern further ---
the base at $T = 161$ for example consists of 12 copies of the cycle
\eq{eq:pattern-12} (with the second $-12$ coming from a blown-up
$-11$), followed by the sequence $\ldots, -12, -1, -2, -2, -3, -1, -5,
-1, -3, -2, -1, -8, -1, -2, -3, -2, -1, -8, 0$ --- the short regular
pattern at the end
with alternating $-8$
and $-2, -3, -2$ clusters corresponds to one of the cyclic linear chains
described in \cite{clusters}.

\section{Weierstrass models}
\label{sec:Weierstrass}

From the toric description we can readily identify the monomials
appearing in the Weierstrass description \eq{eq:Weierstrass} of the
elliptic fibration over any given base.  This provides a tool for
analysis of the class of models over any given base.  It is also
illuminating to match the number of degrees of freedom in the
Weierstrass model to the number of scalar hypermultiplets expected in
the supergravity theory.  In this section we describe the details of
these calculations and clarify some subtleties in the degree of
freedom counting.

\subsection{Monomials in toric bases}

Just as the set of holomorphic functions associated with a given cone
$\sigma$ is described by the dual cone $\sigma^*$, there is a simple
toric description of the monomials corresponding to sections of a
given line bundle in the local coordinate patch associated with each
cone.  To describe the monomials in the Weierstrass functions $f, g$
we need to characterize sections of the line bundles $-4K, -6K$.
Recall that for a toric variety, the anticanonical divisor is given by
the sum of the divisors associated with the vectors $v_i$.  It follows
that for a 2D cone $\sigma$ generated by the vectors $v_1, v_2\in N$, the
condition that a monomial described by a dual vector $m \in M$
is a section of $-n K$ is that $\langle m, v_1 \rangle \geq -n$.
This condition can be geometrically characterized in $M$ by the
condition that $m$ must lie in the cone spanned by a pair of
generators $u_1, u_2$ for the dual cone $\sigma^*$, with the base of
the cone offset by the vector $-n (u_1 + u_2)$ from the origin.

\begin{figure}
\begin{center}
\begin{picture}(200,350)(- 100,- 180)
\put(-90, 100){\makebox(0,0){\includegraphics[width=5cm]{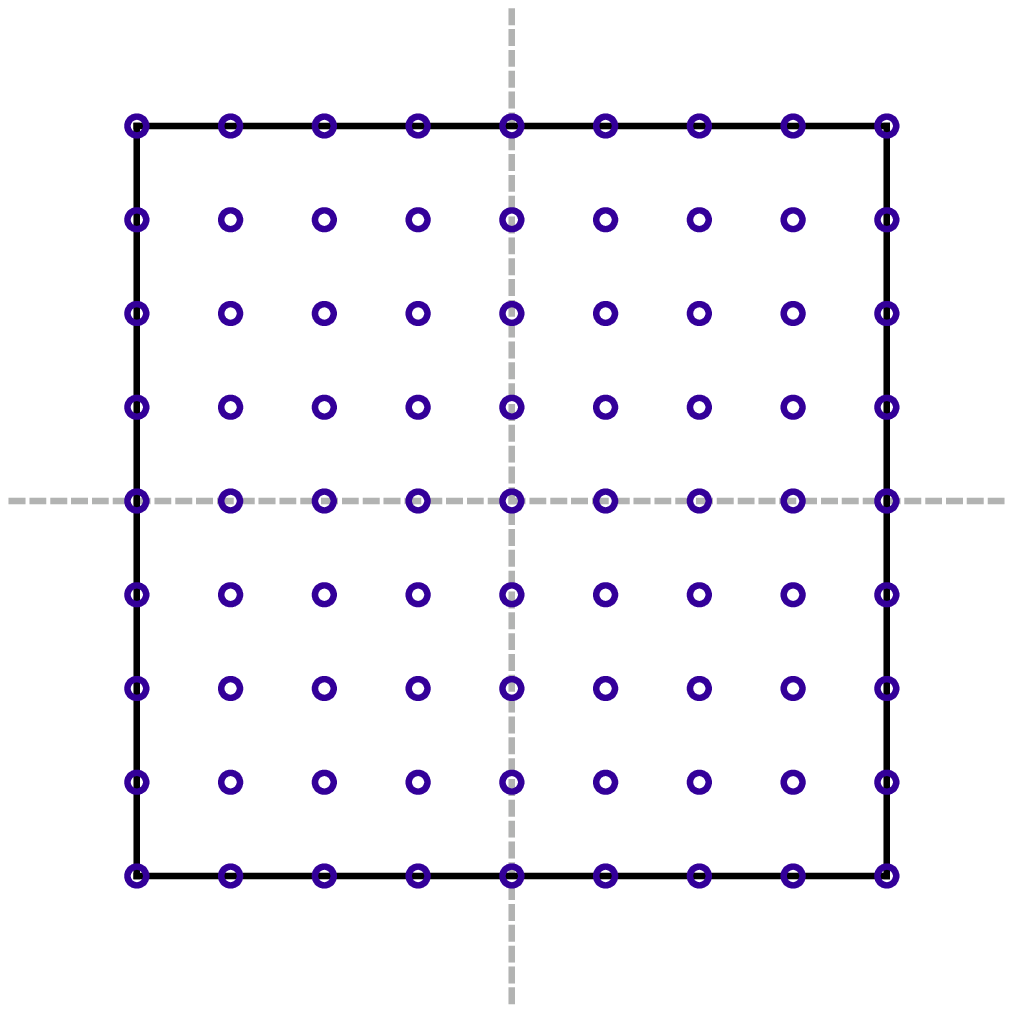}}}
\put(90, 100){\makebox(0,0){\includegraphics[width=7.5cm]{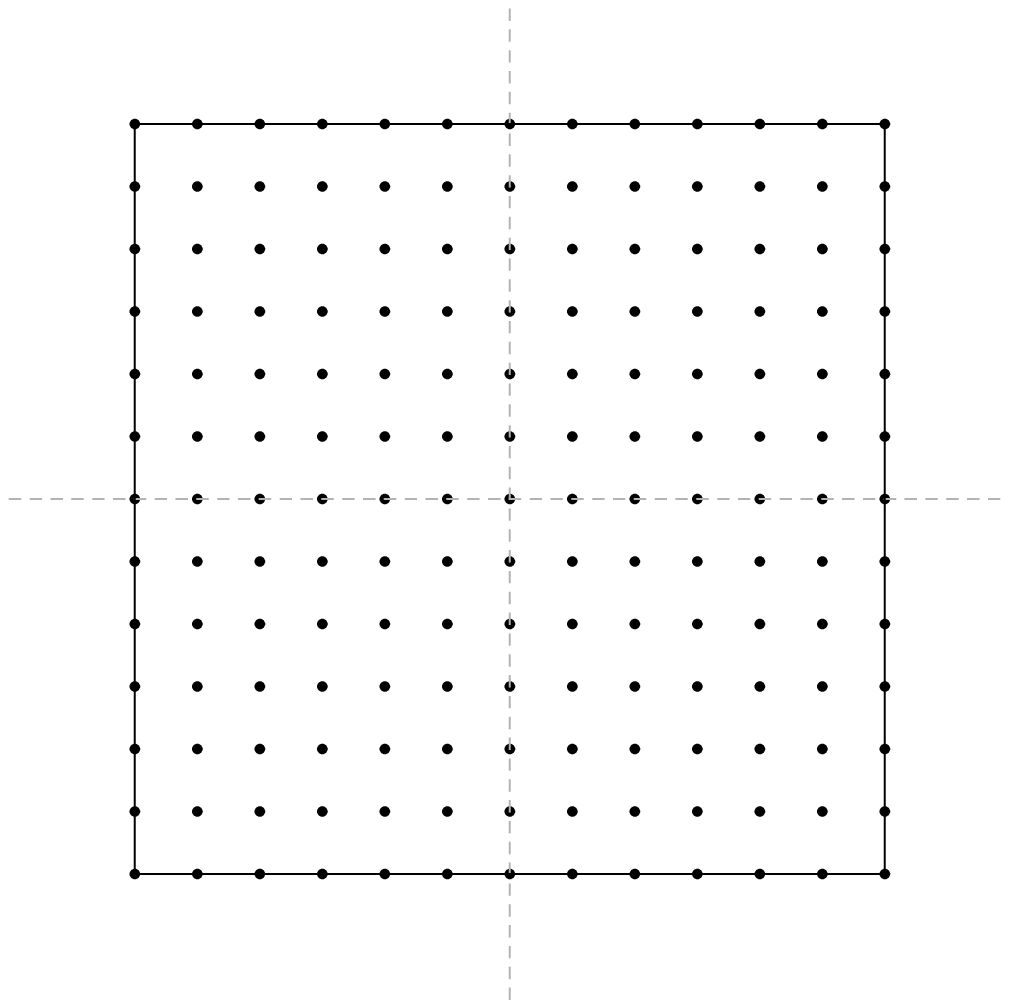}}}
\put(5,-85){\makebox(0,0){\includegraphics[width=14cm]{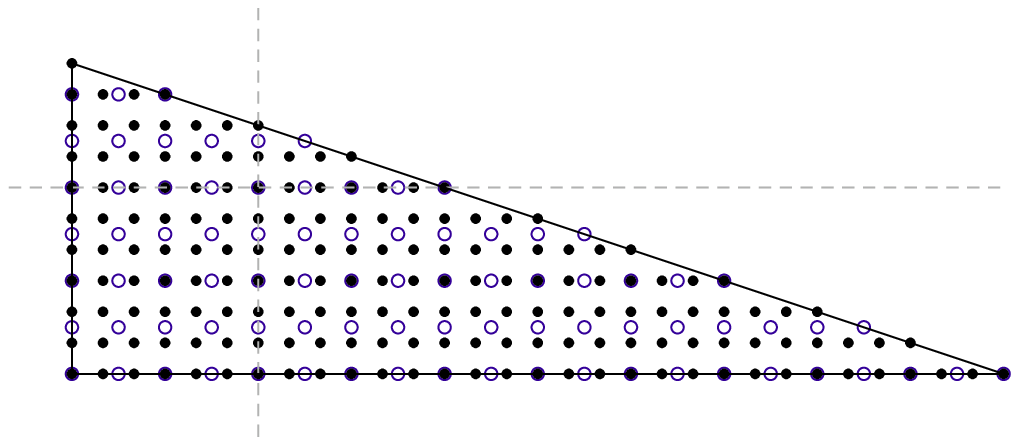}}}
\put(-180,100){\makebox(0,0){$\F_0$}} 
\put(-190,-100){\makebox(0,0){$\F_3$}} 
\put(-100, 30){\makebox(0,0){$f$}}
\put(100, 10){\makebox(0,0){$g$}}
\put(0, -170){\makebox(0,0){$f,g$}}
\end{picture}
\end{center}
\caption[x]{\footnotesize The monomials appearing in $f \in -4K, g \in
  -6K$ for the bases $B=\F_0$  and $\F_3$.  For $\F_0$ the
  monomials in $f$ and $g$ are shown separately.  Since the geometry
  of the bounding region is identical in both cases up to scale, for
   $\F_3$ the monomials are superimposed, with open (colored) circles
  indicating monomials in $f$ and small closed (black)
  circles indicating monomials in $g$. 
Large (multi-colored) circles indicate monomials in both $f$ and $g$.}
\label{f:simple-monomials}
\end{figure}

Based on this characterization, we can easily construct the set of
monomials in the global Weierstrass description of any toric base.
The monomials for the minimal bases $\F_0$ and $\F_3$ are shown in
Figure~\ref{f:simple-monomials}.
For $\F_m, m \geq 2$ the general pattern is that $f$ is constrained by a right
triangle with base containing $4 (m + 2) +1$  points with a slope of
$1/m$ on the diagonal, while $g$ is constrained by a similar triangle
with
$6 (m + 2) + 1$ points on the base.  The total number of monomials for
$\F_m, m \geq 2$ is then
\begin{equation}
W = (2 m   + \frac{9 + (9 \; {\rm mod} \; m)}{2})
(5 +\lfloor \frac{9}{m}\rfloor )+
 (3 m  + \frac{13 + (13 \; {\rm mod} \; m)}{2})
(7 +\lfloor \frac{13}{m}\rfloor )\,.
\end{equation}
This number is computed for each $m$ and tabulated in Table~\ref{t:DOF}.

Note that from the available monomials we can immediately check that
the degrees of vanishing of $f$ and $g$ on each divisor are those
expected.  For example, for the $-3$ curve on $\F_3$, associated with
the vector $v_3 = (0, -1) \in N$,
it is clear from
Figure~\ref{f:simple-monomials} that there are no monomials $m$ with $
\langle m, v_3 \rangle = -4 + k$ for $k = 0, 1$, so that $f$ vanishes
to order $2$ (but not higher) on that divisor.

\subsection{Degrees of freedom and the gravitational anomaly
  constraint}

The spectrum of any consistent 6D supergravity theory must obey the
gravitational anomaly constraint \cite{gsw, Sagnotti, Erler}
\begin{equation}
H-V= 273-29T \,,
\label{eq:constraint}
\end{equation}
where $H, V,$ and $T$ are the numbers of massless scalar
hypermultiplets, vector multiplets, and tensor multiplets in the
theory.  The physical degrees of freedom in the Weierstrass
coefficients $f, g$ correspond to massless neutral hypermultiplets in
the maximally Higgsed 6D theory.  Thus, there is a close
correspondence between the monomials in $f, g$ determined from the
toric data and the spectrum of the theory.  Matching these degrees of
freedom for the generic model over toric bases gives insight into a
number of features of the theory.  

In general, there is a redundancy in the Weierstrass parameters
associated with the dimension $w_{\rm aut}$ of the automorphism group
of the surface $B$.  For $\F_m,$ $w_{\rm aut}=m + 5$ when $m > 0$, and
$w_{\rm aut} = 6$ for $\F_0$ (see Table~\ref{t:DOF}).
There is additionally an overall scale factor
that can be removed from $f, g$ without changing the geometry.
Furthermore, in all cases one scalar field is not a deformation in $f,
g$ but rather arises from the overall K\"ahler modulus of the F-theory
base.  Finally, as discussed further below, curves of
self-intersection $-2$ that do not live in clusters carrying a gauge group arise at
codimension one points in the moduli space, so the degree of freedom
removing these curves does not appear in the Weierstrass degrees of
freedom for a given base.  Putting these contributions together, the number
of Weierstrass monomials is related to the number of hypermultiplets
in a maximally Higgsed model over any F-theory base through
\begin{eqnarray}
H_{\rm neutral}  =H-H_{\rm charged} 
 & = &W  -w_{\rm aut}+ N_{-2}-w_{\rm scaling}  + h_{\rm Kaehler}   \nonumber\\
 & = & W  -w_{\rm aut}+ N_{-2}\,,\label{eq:hw1}
\end{eqnarray}
where $N_{-2}$ is the number of $-2$ curves on the base $B$ that are not in a cluster carrying a gauge group, and $H_{\rm charged}$ is the number of
non-Higgsable charged matter fields in the generic model over that
base.

In looking at different toric models it is helpful to decompose the
extra Weierstrass degrees of freedom coming from the automorphism
group of the base into contributions from the different divisors.  
In general, there are $k +1$ redundant Weierstrass degrees of
freedom associated with a divisor of self-intersection $k \geq 0$ in
the toric fan.  This is easily seen from the description of
the automorphism group given by Cox \cite{cox-homogeneous}, 
in terms
of the homogeneous coordinate ring of the toric surface and the
associated ``polar polytope''.  There is 
a simple combinatorial description of the  polar polytope:
it is the set of vectors ${m}\in M$
that satisfy $\langle m , n \rangle  \geq -1$ for all of the generators
$n$ of the fan of the toric surface.  This polytope is similar to the
regions of points graphed in Figure~\ref{f:simple-monomials}, but with
a rescaling of the lattice.  Each lattice point which is in the
interior of a codimension one face of the polar polytope is associated
with a one-parameter subgroup of the automorphism group, and these
are all independent.  (See also \cite{Aspinwall:1993nu}).  For a curve of self-intersection
$k$, since $-K_S\cdot C = k+2$, the corresponding edge of the polar polytope
has $k+3$ lattice points, $k+1$ of which are interior lattice points.

For example, for $\F_0$, each curve associated with a
divisor $v_i$ in the toric fan has self-intersection 0.  Each such
curve has a single degree of freedom associated with fixing the
position of the curve in the given toric coordinates (for $D_1, D_3
\sim F$ this corresponds to fixing a coordinate $z = z_0$, and for
$D_2, D_4$ this corresponds to fixing a coordinate $w = w_0$).  These
degrees of freedom are removed when we blow up points on these curves
that fix the locations of the proper transforms in the new base $B'$.
There are also two Weierstrass degrees of freedom present for every
toric variety that cannot be removed by blowing up, associated with
rescaling of the two toric coordinates $z, w$.
Combining these contributions with the missing degrees of freedom for
$-2$ curves,
we
write the total number of ``extra'' Weierstrass degrees of freedom as
\begin{equation}
w_{\rm extra} =\sum_{k \geq -2}(k +1) N_k + 2 
=W-H_{\rm neutral}\,.
\label{eq:hw2}
\end{equation}
The dimension of the automorphism group is the sum of all these
contributions except for the missing moduli from $-2$ curves
\begin{equation}
w_{\rm aut} = w_{\rm extra} + N_{-2} \,.
\end{equation}

To see how this correspondence works in an example, consider the
simple case of $\F_0=\P^1 \times \P^1$.  In this case we have (see
Figure~\ref{f:simple-monomials})
\begin{eqnarray}
f  =   \sum_{a = -4}^{4} \sum_{b = 4}^{4}f_{ab} z^aw^b ,  &\;\;\;\;\; &
g  =   \sum_{a = -6}^{6} \sum_{b = 6}^{6}g_{ab}z^aw^b \,.
\end{eqnarray}
Thus, there are $W =81 + 169 = 250$ coefficients $f_{ab}, g_{ab}$ associated with
independent monomials in the Weierstrass parameters $f, g$ for $B =
\F_0$.  Comparing to \eq{eq:constraint}, the generic model over this
base has $T = 1$ and a completely Higgsed gauge group ($V= 0$), so we
expect $H = 244$.  This matches with \eq{eq:hw1} since $w_{\rm aut} =
6$ for $\F_0$.

Another useful case to consider is $\F_2$.
As mentioned above,
part of the difference between the number of Weierstrass moduli
and the number of hypermultiplets arises from curves of
self-intersection $-2$.
In general, a curve of self-intersection $-2$ that does not connect to
a cluster containing
other curves of self-intersection $-3$ or less (and that therefore
does not carry a non-Higgsable gauge group factor by
Table~\ref{t:clusters}) arises in the branch of moduli space
associated with a base without that $-2$ curve at a locus of complex
codimension one.  For example, $\F_2$, which is topologically
equivalent to $ \P^1 \times \P^1$, arises on a codimension one locus
in the space of $\F_0$.  \footnote{This is seen by embedding $\F_0$ into $\P^3$
as a nonsingular hypersurface of degree two, and taking a limit to 
a singular hypersurface (which is isomorphic to $\F_2$ with the
$-2$ curve blown down).  The singularities of that family of surfaces
can be resolved simultaneously, giving a family linking $\F_0$ and
$\F_2$.}
Thus, we  associate with each curve of self-intersection
$-2$ a deficit in the number of degrees of freedom counted by
Weierstrass moduli of $ -1$, as in \eq{eq:hw1}.  For $\F_2$, the
number of Weierstrass monomials is 250, and the automorphism group has
dimension 7, so only 243 degrees of freedom (including the overall
scaling factor that compensates for the missing K\"ahler modulus) are
described by the Weierstrass monomials.  The remaining scalar field
that brings the total to 244 is the one that was tuned to produce the
$-2$ curves.

For each of the Hirzebruch surfaces $\F_m$, it is straightforward to
verify that the counting of degrees of freedom matches both with the
size of the automorphism group through \eq{eq:hw1} and with the
degrees of freedom associated with specific divisors through
\eq{eq:hw2}.  In general, there are $m + 5$ extra degrees of freedom
in $W$, associated with $w_{\rm aut}$, or $m +1$ from the divisor of
self-intersection $m$, 2 from the two $0$-curves, and 2 universal
extra degrees of freedom.  Note that for $m = 7$, $W-w_{\rm aut} =
W-w_{\rm extra} = 349 = H-28$.  This counting indicates that the 28
non-Higgsable hypermultiplets (56 half-hypermultiplets) charged under
$\gf_4$ in the 6D supergravity theory associated with this F-theory
compactification do not appear as parameters in the Weierstrass model.
This makes sense as only neutral scalars should appear as moduli in
the Weierstrass form.

\begin{table}
\centering
\begin{tabular}{|c|c|c|c|c|c|c|c|c|c|c|
}
\hline
$m$ & 0 & 1 & 2 & 3 & 4 & 5 & 6 & 7 & 8 
& 12 \\
\hline
$H = 244 + V$ & 244& 244 & 244 & 252 & 272& 296 & 322& 377 & 377 & 492\\
$W$ & 250 & 250 & 
250 & 260 & 281 & 306 & 333 & 361 & 390
& 509\\
$w_{\rm aut}$ & 6 & 6 & 7 & 8 & 9 & 10 & 11 & 12 & 13 
 & 17\\
\hline
\end{tabular}
\caption{Matching of degrees of freedom between Weierstrass
  coefficients computed using toric data and massless field content
  for F-theory compactifications over bases $\F_m$.  $H = 244 + V$ is
  the number of scalar hypermultiplets.  $W$ is the number of scalar
  coefficients in the Weierstrass functions $f, g$.  $w_{\rm aut}$ is
  the dimension of the automorphism group of the base.  In each case,
  $H -H_{\rm charged} = W-w_{\rm aut}+ N_{-2}$, where $N_{-2}$ is the
  number of $-2$ curves not carrying a gauge group factor.  $w_{\rm
    aut}$ can be decomposed into contributions of $m +1$ from each
  irreducible effective curve of self-intersection $m \geq 0$, plus 2
  universal redundant degrees of freedom} \label{t:DOF}
\end{table}

\subsection{Blowing up points and degrees of freedom}

Having checked that the degree of freedom counting matches for the
Hirzebruch surfaces $\F_m$, we can perform a sequence of successive
toric blow-ups and match the decrease in degrees of freedom to changes
in the divisor structure.  From the gravitational anomaly constraint
(\ref{eq:constraint}) or from simple geometric considerations on the
local form of the Weierstrass model, we know that generally 29 scalar
degrees of freedom are removed when a point is blown up.  The number
of Weierstrass degrees of freedom lost may differ if curves of
self-intersection $-2$ or $k \geq 0$ are involved, due to the under-
or over-counting associated with these divisors.  Furthermore, if the
non-Higgsable gauge group structure is changed then the number of
massless scalars will change accordingly, in correspondence with the
gravitational anomaly constraint (\ref{eq:constraint}).  The change in
degrees of freedom from the different kinds of blow-ups, for example
blowing up a point between  the $-2$ curve and a $-3$ curve in the
sequence $-1, -2,
-3, -1$, can in principle be checked by doing a local analysis of the
toric geometry using only the nearby relevant divisors.  We have done
the check directly on the global models, by systematically
verifying for all
61,539 toric bases that the matching between Weierstrass parameters
and neutral scalar fields works out.  
This confirms the expectation that in general global
models the gauge group and matter content on any non-Higgsable cluster
of intersecting divisors is the minimum required from the intersection
of the divisors with $-K$.
As a particular example, the largest base with $T = 193$ has a
Weierstrass model with 15 monomials, encoding 12 physical degrees of freedom.

\begin{figure}
\begin{center}
\begin{picture}(200,260)(- 100,- 100)
\put(80,110){\makebox(0,0){\includegraphics[width=6cm]{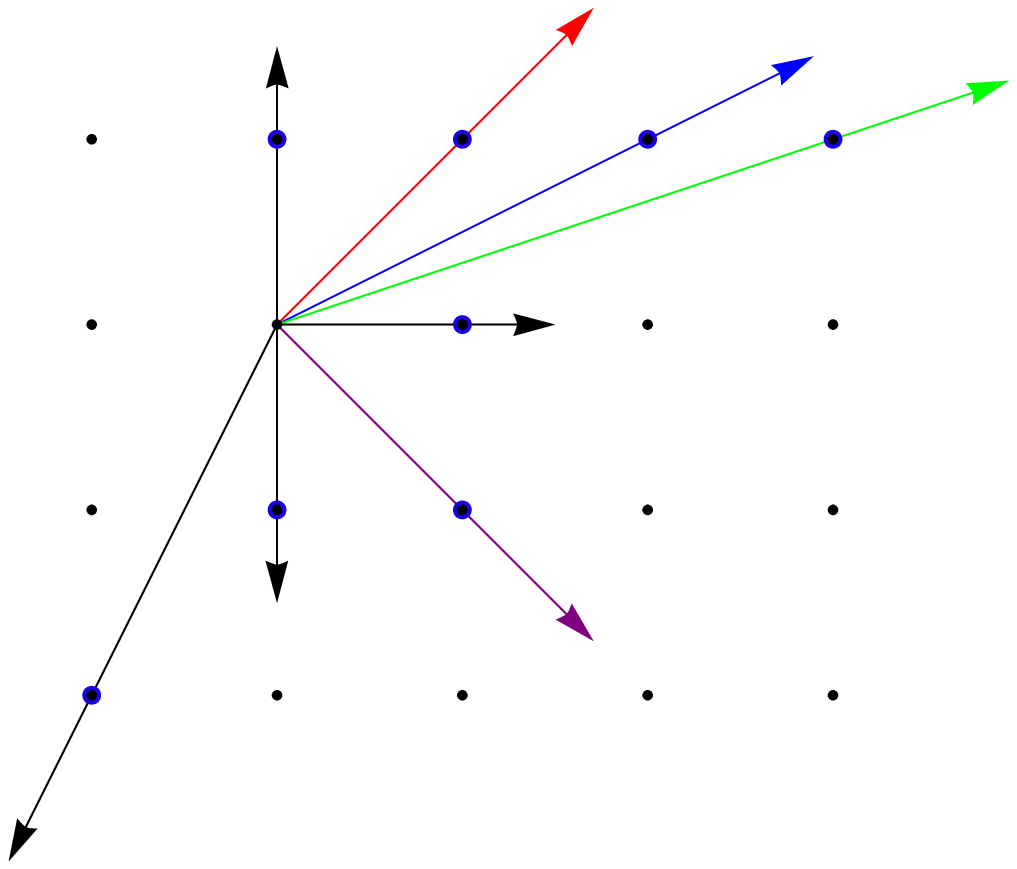}}}
\put(-60,0){\makebox(0,0){\includegraphics[width=12cm]{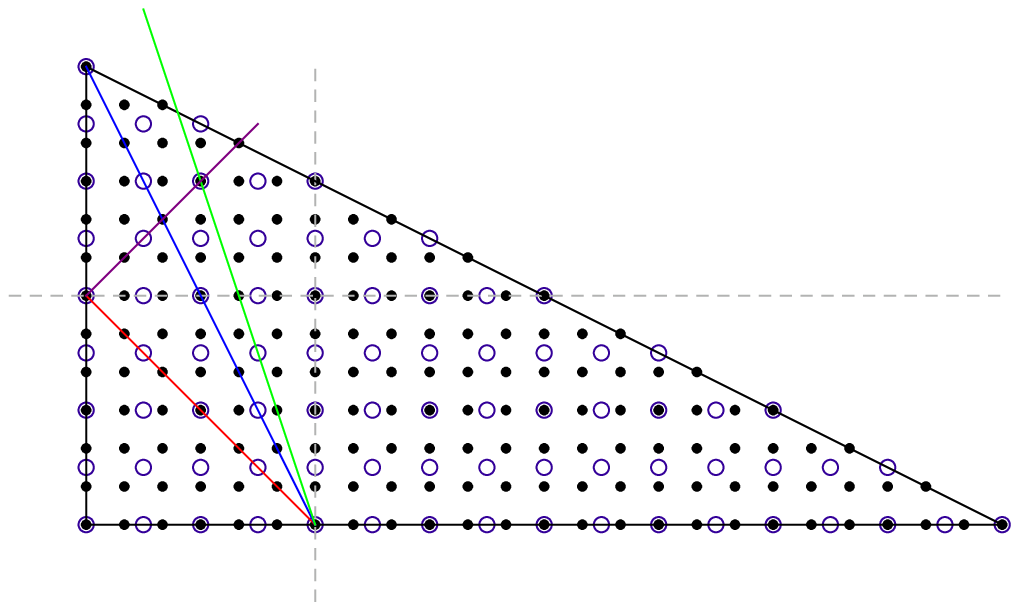}}}
\put(55,40){\makebox(0,0){$N$}}
\put(-90,-90){\makebox(0,0){$M$}}
\end{picture}
\end{center}
\caption[x]{\footnotesize  An example of removal of Weierstrass
  degrees of freedom when points on the base are blown up.  
Starting from $\F_2$,
the first
  blow-up ($(1, 1)$ in the toric fan) removes 29 massless scalar
  hypermultiplets but 31 Weierstrass moduli.  Details of the
  correspondence between hypermultiplet and Weierstrass moduli
  counting are in the text.}
\label{f:counting-example}
\end{figure}

It may be instructive to follow an explicit example through a sequence
of blow-up operations and confirm that the number of Weierstrass
degrees of freedom varies as expected.
Figure~\ref{f:counting-example} depicts a sequence of four blow-up
operations on $\F_2$.  
In the toric fan, the following sequence of points is blown up: $(1,
1)$ (red), $(2, 1)$ (blue), $(3, 1)$ (green), $(1, -1)$ (purple).
In the resulting sequence of bases $B_1 = \F_2,
B_2, \ldots, B_5$, we have the
following structure for the gauge algebra and number of Weierstrass
degrees of freedom, where for each base $I = (D_1 \cdot D_1, \ldots D_{k} \cdot
D_k)$ indicates the self-intersection numbers of the divisors (always
starting with the toric divisor associated with $v= (0, 1)$, and with
the exceptional divisor at each stage noted in bold)
\begin{align}
&I_1 = (2, 0, -2, 0) & &  \G_1 = 0 & & W = 250\\
&I_2 = (1, {\bf -1}, -1, -2, 0) & &  \G_2 = 0 & &  W = 219\\
&I_3 = (1,  -2, {\bf -1}, -2, -2, 0) & &  \G_3 = 0 & &  W = 188\\
&I_4 = (1,  -2, -2, {\bf -1}, -3, -2, 0) & &  \G_4 = \gg_2 \oplus
\gsu (2) & &  W = 169\label{eq:gs}\\
&I_5  = (1,  -2, -2, -1, -4, {\bf -1}, -3, 0) & &  \G_5 = 
\gso (8) \oplus
\gsu (3) & &  W = 167\,.
\end{align}
This sequence of blow-up operations and the resulting constraints on
the monomials $m \in M$ are shown geometrically in
Figure~\ref{f:counting-example}.  In the first step, 31 Weierstrass
degrees of freedom are removed.  This can be understood as the 29
physical scalars that are traded for the extra tensor, plus two
unphysical Weierstrass moduli associated with the first two divisors
in $I_1$, whose self intersections decrease from $2$ and $0$ to $1$
and $-1$ ({\it i.e.}, $w_{\rm extra}$ is reduced from 4 to 2).  At the
next stage again 31 Weierstrass degrees of freedom are removed.  29 of
these again are physical scalars, and two are the extra moduli
associated with fixing $-2$ curves in $B_3$ that do not support a gauge group
($N_{-2}$ goes from 1 to 3).  For the base $B_4$, the model develops a
non-Higgsable $(-3, -2)$ cluster that requires a gauge group of $\gg_2
\oplus \gsu (2)$.  For this model we have $T = 4$, a gauge group of
dimension $V=17$, and 8 charged hypermultiplets.  The total number of
scalar hypermultiplets should be $H = 273 + V-29T =174$.  There are 2
universal extra Weierstrass parameters, $w_{\rm extra} = 1$ (from $2+1$
extra Weierstrass parameters for the curves of self-intersection $1$
and $0$, and two missing moduli from the two $-2$ curves that do not
carry gauge groups), so the number of moduli from the Weierstrass
counting should be $169-3 = 166$.  The maximally Higgsed model over
this base has 8 charged scalar hypermultiplets (16
half-hypermultiplets) that cannot be Higgsed, from
Table~\ref{t:clusters}.  From this counting it is clear that, just as
for the matter charged under $\gf_4$ on a $-4$ curve, these
non-Higgsable charged matter fields do not appear in the Weierstrass
parameters for this model.  In the final model with base $B_5$, we
have $T = 5$, a gauge group of dimension $V=36$, and no matter, so we
expect $H = 273 + V-29T = 164$.  This is in exact agreement with $W =
167$ since there are 2 universal extra Weierstrass parameters, and
$w_{\rm extra} = 1$ (from $2+1$ extra Weierstrass parameters for the
curves of self-intersection $1$ and $0$, and two missing moduli from
the two $-2$ curves that do not carry gauge groups).  We can
furthermore check that in this sequence of models the degrees of
vanishing of $f, g$ on the various curves are precisely the minimal
values expected.  In particular, we can check that the degrees of
vanishing of $f, g$ on the $-3, -2$ curves in $I_4$ are $(2, 3)$ and
$(1, 2)$, as computed in \cite{clusters}, by confirming for example
that all monomials $m$ in $f$ satisfy $\langle m, D_{-3} \rangle \geq
2$, etc.

As discussed above in the case of $\F_m$ and the blown-up base $B_3$
from \eq{eq:gs}, charged hypermultiplets associated with non-Higgsable
matter content do not appear as Weierstrass degrees of freedom.  This
makes sense, as Weierstrass degrees of freedom are generally uncharged
scalar moduli (with the exception of the redundant unphysical degrees
of freedom discussed above).  This raises the natural question of
where the moduli associated with the non-Higgsable charged
hypermultiplets arise in the F-theory description.  Geometrically,
from the M-theory point of view the charged matter corresponds to
additional 2-cycles in the resolved geometry.  
Like the cycles
associated with
gauge fields
on the 7-branes, however, these M-theory degrees of freedom are not included in
the geometrical degrees of freedom in the Weierstrass description of
an F-theory model.  Another example of this appears
when the Weierstrass parameters are tuned to produce an
enhanced gauge symmetry on some particular divisor. As discussed in
detail in \cite{Morrison-Taylor} in the case of the $\P^2$ base, and
in \cite{KMT} for the bases $\F_m,$ $m = 0, 2$, the number of moduli
that are tuned to produce, for example, an $SU(N)$ gauge group is
given by the dimension of the group, $N^2 -1$.  Thus, at an enhanced
symmetry point, the Weierstrass parameters that are fixed become some
of the matter fields that are charged under the enhanced gauge group.
The total number of fields also increases, however, at the enhanced
symmetry locus by the dimension of the group.  Thus, extra charged
matter fields must be included when degree of vanishing of the divisor
locus is enhanced on some particular curve.  Physically, this
corresponds to extra degrees of freedom occurring when 7-branes become
coincident.  On coincident 7-branes there are indeed extra fields,
scalar fields that live in the adjoint of the gauge group.  These are
the fields used in \cite{T-branes} to describe additional structure of
the Hitchin bundle over the branes in the structure referred to as
``T-branes''.  It is natural to speculate that in all cases, the
additional ``missing'' charged scalar fields in the Weierstrass model
correspond to such fields on the branes.  In fact, it should also be
noted that the 6D supergravity scalar hypermultiplet fields are
quaternionic and each contain four real degrees of freedom.  The
Weierstrass moduli are complex and only contain two degrees of
freedom; the remaining degrees of freedom are contained within the
7-branes and not usually treated within the F-theory context.
Associating the extra charged matter fields with fields on the brane
as just suggested would naturally complete the counting of degrees of
freedom, although a more explicit description of the non-Higgsable
charged matter fields in this fashion is desirable.  In any case, the
usual formulation of F-theory in terms of Weierstrass models does not
contain a number of the degrees of freedom of the theory, such as the
overall K\"ahler modulus and these charged moduli and the other gauge
degrees of freedom on the 7-branes, that in the 4D context can carry
fluxes.  At this stage, while F-theory is a useful way of
characterizing a general class of nonperturbative string vacua, there
is still no systematic formulation that gives a complete description
of the corresponding low-energy supergravity theory.  Some recent work
on a more complete correspondence through M-theory appears in \cite{Bonetti-Grimm}

\section{Further directions}
\label{sec:further}

\subsection{Non-toric bases}
\label{sec:non-toric}

A natural extension of this work is to attempt a systematic
classification of all F-theory bases without the toric constraint.  It
has been proven that the number of birational equivalence classes of
elliptically fibered Calabi-Yau threefolds is finite \cite{Grassi91,Gross}.  A
simple argument based on the Weierstrass parameterization shows that
there are a finite number of branches of the F-theory moduli space
associated with distinct bases \cite{KMT-II}.  But there is as yet no
systematic classification of all such bases or elliptic fibrations.
To extend the classification beyond the toric set, more complicated
intersection structures of divisors must be considered.  As analyzed
in \cite{clusters}, there are strong constraints on what types of
intersections can arise.  The only ways in which a set of curves of
self-intersection $k \leq -2$ can intersect one another are those
tabulated in Table~\ref{t:clusters}, and all ways in which a $-1$
curve can intersect these clusters are listed in \cite{clusters}.
Nonetheless, the combinatorial possibilities become large as the
number of tensor multiplets increases, so a systematic classification
becomes challenging.  One approach is to begin with a bound on $T$.
While we believe that the (essentially) toric example described here
and in \cite{Candelas-pr, Aspinwall-Morrison-instantons} with $T =
193$ is the base with the largest value of $T$, this has not been proven.
We describe here briefly two approaches that could be taken to bound $T$
and/or to systematically classify more general F-theory bases.  We
leave further elaboration of these ideas to  future work.

For toric bases the irreducible effective curves of negative
self-intersection form a simple closed loop of pairwise intersections.
The main new features that appear in considering non-toric bases are
branches, where one curve can intersect 3 or more other curves, and
additional loops.
We can construct a large family of non-toric bases that have branching
and additional loops but that still have enough of the structure of
toric surfaces that they can be studied in a controlled fashion.  We
do this by considering a slightly more general class of blow-ups of
the Hirzebruch surfaces $\F_m$.  As discussed in the Appendix, the
toric surfaces for F-theory can all be constructed by blowing up only
the points at which a pair of fibers $F_1, F_2$ intersect the sections
$S_\infty, S_0$ of self-intersection $\pm m$, or after blowing up such
points, further blowing up at intersection points between pairs of
curves of negative self-intersection arising from the blow-up of the
$F_i$.  We can explore a larger space of bases by simply including
more fibers but only blowing up points in the same fashion.  In
particular, we do not blow up any points that are on a fiber $F$ but
not at an intersection  either between $F$ and $S_\infty$ or $S_0$, or
between two curves of negative self-intersection that lie within $F$.
This choice preserves the $\C^*$ action on the fibers that is part of
the toric $(\C^*)^2$ action.
In general, it is difficult to tell which new curves of negative
self-intersection will appear after blowing up a sequence of points,
but in this case with a residual $\C^*$ action no new curves need to
be added other than the exceptional curves from the blow-ups.

We can in principle systematically describe all non-toric bases with
the structure just described.  We report here on only one simple
experiment in this direction, which provides supporting evidence for
the conjecture that there are no models with $T > 193$.  We have
considered possible non-toric bases that can be formed in the class
described above by blow-ups on three distinct fibers $F_1, F_2, F_3$.
We start from $\F_m$ and blow up the intersections between the $F_i$
and $S_0$ by $k, l, r$ times respectively, with each of $k, l, r
\geq 1,$ and $k + l + r \leq 2m$.  This gives us a ``frame'' similar
to those described in the Appendix but with 3 chains connecting the
divisors $S_\infty, S_0$.  This gives some tens of thousands of
non-toric models, of which the largest has $T = 170$.  The model with
$T = 170$ is produced by choosing $m = 12, k = l = 1, r = 22$, and has
simple chains $(-1, -1)$ replacing $F_1, F_2$, and a chain of length
167 for $F_3$ containing 14 copies of the pattern
(\ref{eq:pattern-12}) without the terminal $-12$'s, and with the 12th
curve from each end having self-intersection $-11$ instead of $-12$.
This base is closely related to the structure of the base with the
largest $T = 193$, and the other bases of this non-toric type with
large $T$ are similar.  The next-largest has $T = 159$, with $F_2$ or
$F_1$ replaced by $(-1, -2, -1)$ and only 13 copies of the $E_8$
pattern in $F_3$.
This analysis indicates that there are no local types of branching
where one curve connects to three that increase the range of possible
$T$'s.  We can understand this geometrically by noting that large $T$
is always associated with chains locally dominated by the pattern
(\ref{eq:pattern-12}).  This follows from the bounds on $T$ for any
given gauge group studied in \cite{clusters}.  From that analysis, and
from the gravitational anomaly bound (\ref{eq:constraint}), it is
clear that increasing $T$ substantially is only possible when the
gauge group is large, and $E_8$ and the associated periodic chain sequence
provides the largest gauge group for the smallest change in $T$.  To
do better than the linear sequence of 16 copies of
(\ref{eq:pattern-12}) with terminal $-12$'s that forms the basis of
the toric model with $T = 193$, somehow the $-12$ curves carrying the
$E_8$ factors would need to be connected in a way that increases the
possible $T$ while still giving a surface that blows down to a
Hirzebruch surface with only one curve of negative self-intersection.
From the connectivity rules in Table~\ref{t:connections}, a $-12$
curve can only be connected by a $-1$ curve to a sequence of curves
of self-intersection $-2, -2, -3$, and this can in turn be connected
to a curve of self-intersection at most $-5$.  This severely limits
the range of possible structures incorporating $-12$ curves.  We
cannot, for example, have three $-12, -1, -2, -1, -3, -1$ chains
attached to a single $-5$ curve, or as we blow down the chains the $-5$ curve
would become a curve of positive self-intersection leaving three
disconnected components.  This is incompatible with the minimal model
result that the surface must blow down to a surface with at most one
curve of negative self-intersection, unless the original three chains
are connected through other $-1$ curves.  In the latter case the
original $-5$ curve plays no role and can be dropped from the
analysis.  While this is a somewhat heuristic analysis, it gives the
flavor of the underlying geometric reason why it is difficult to
connect $-12$ curves in any way that can allow $T > 193$.  Combining
this with the systematic analysis of non-toric models with three
fibers described above seems to give  strong evidence that  $T = 193$
is indeed the largest value of $T$ for any 6D F-theory model, even
allowing non-toric structure for the base.  A rigorous proof of this
result would, however, be  nice to have.

While the preceding general arguments suggest that the upper bound on
$T$ is indeed 193, a complete proof of this statement may be somewhat
nontrivial due to the combinatorial complexity of the set of possible
combinations of intersecting divisors when the toric constraint is
relaxed.  The number of possible distinct ways that a $-1$ curve can
intersect a combination of non-Higgsable clusters is 183, as shown in
\cite{clusters}.  Considering combinations with more than a few
clusters leads to a rapid combinatorial growth in possibilities.
Nonetheless, using some general principles it may be possible to put a
fairly strict bound on $T$, and possibly even to prove the bound $T
\leq 193$.  To illustrate how this might work we briefly consider a
simplified piece of the problem.  
The remainder of this subsection is not relevant for the rest of the
paper, and the reader not interested in worrying about how to bound
$T$ for non-toric bases can skip to Section \ref{sec:constraints}.

\begin{figure}
\begin{center}
\begin{picture}(200,170)(- 100,- 90)
\put(-120,-80){\makebox(0,0){(a)}}
\put(0,-80){\makebox(0,0){(b)}}
\put(120,-80){\makebox(0,0){(c)}}
\put(-120, 40){\circle{30}}
\put(-120, 55){\circle*{5}}
\put(-120, 25){\circle*{5}}
\put(-105, 40){\circle*{5}}
\put(-135, 40){\circle*{5}}
\thicklines
\put(- 135, -25){\line( 0, -1){40}}
\put(- 105, -25){\line( 0, -1){40}}
\put(-140,-30){\line(1, 0){40}}
\put(-140,-60){\line(1, 0){40}}
\thinlines
\put(-120, 15){\vector( 0, -1){30}}
\put(120, 15){\vector( 0, -1){30}}
\put(0, 15){\vector( 0, -1){30}}
\put(20, 20){\circle*{5}}
\put(-20, 20){\circle*{5}}
\put(0, 65){\circle*{5}}
\put(0, 40){\circle*{5}}
\put(0,40){\line( 0,1){ 25}}
\put(0,40){\line( 1, -1){20}}
\put(0,40){\line( -1, -1){20}}
\thicklines
\put(-20,-45){\line(1, 0){40}}
\put(-15,-60){\line(1, 1){30}}
\put(15,-60){\line(-1, 1){30}}
\thinlines
\put(132,57){\circle*{ 5}}
\put(132,33){\circle*{ 5}}
\put(120,45){\circle*{ 5}}
\put(100,45){\circle*{ 5}}
\put(132,45){\circle{24}}
\put(100,45){\line(1, 0){20}}
\thicklines
\put(100,-45){\line(1, 0){40}}
\end{picture}
\end{center}
\caption[x]{\footnotesize Configurations of $-4$ curves connected by
  $-1$ curves (depicted as graphs with $-4$ curves as nodes), and the
  configurations of $-2$ curves (depicted as bold lines) reached after
  blowing down all $-1$ curves needed to reach a generalized del Pezzo
  surface}
\label{f:4-configurations}
\end{figure}
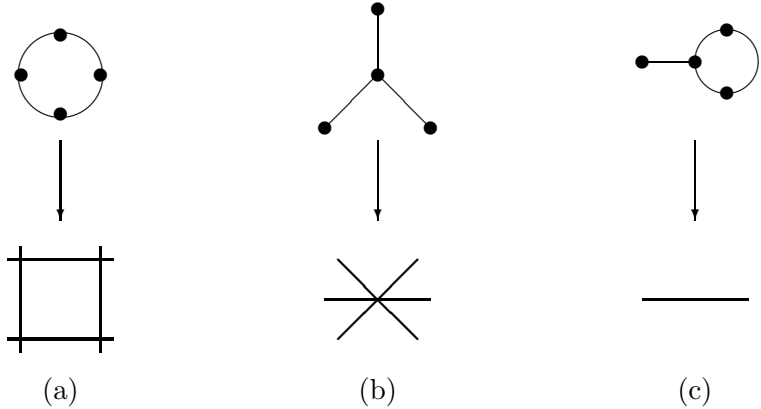

We consider the subclass of F-theory
bases that contain only $-1$ and $-4$ curves.  This corresponds to
low-energy supergravity theories with a gauge group $SO(8)^k/\Gamma$
and no matter in the maximally Higgsed phase, where $\Gamma$ is a
discrete quotient that will not concern us.  From the general
arguments in \cite{clusters} and the value of $\Delta T = 1$ in
Table~\ref{t:clusters}, we know that any theory with a maximally
Higgsed gauge group containing only $k$ $\gso(8)$ summands has a
maximum value of $T \leq 9 + k$.
There are only a few ways in which $-4$ curves can be connected by
$-1$ curves.  From the Table in \cite{clusters}, a $-1$ curve can only
(i) intersect a single $-4$ curve once, (ii) intersect two distinct
$-4$ curves once each, or  (iii) intersect a single $-4$ curve twice.
This limits the range of possibilities, and makes a general argument
bounding both $k$ and $T$ possible without too many combinatorial
complications.  The basic idea is to  consider all possible ways in
which a combination of $-4$ curves can be connected and blown down to
a generalized del Pezzo surface containing no curves of
self-intersection $-3$ or below.  Note that any surface with $T > 3$
with no curves of self-intersection $-5$ or less
that can be blown down to $\F_4$ can also be blown down to $\F_2$ ---
since the points blown up on the $\F_4$ must lie off the $-4$ curve
--- and therefore can be blown down to a generalized del Pezzo.
Generalized del Pezzo surfaces have
at most $T = 9$.  For $T < 9$, a generalized del Pezzo surface has at
most $T$ $(-2)$-curves, and for $T = 9$ the maximum number of such
curves is $12$.  The generalized del Pezzo surfaces at $T = 9$ are
\new{rational elliptic surfaces}, for which a complete list is known
of the possible intersection configurations of $-2$ curves
\cite{Persson, Miranda}.\footnote{Technically, those papers only consider
the cases of rational elliptic surfaces with section, and one must
also consider the ``Halphen pencil'' cases as well \cite{dolgachev-halphen}.
But since the Jacobian fibration of a Halphen pencil is a rational
elliptic surface with a section, the configuration of non-multiple
singular fibers does not change.}
While the number of $-1$ curves becomes
infinite at $T = 9$, we need only consider a finite subset.  For any
valid F-theory base containing a
realizable configuration of $-4$ curves there is at least one finite
set of $-1$ curves that can be added to the $-4$ curves forming a
network with the property that all curves needed to blow down to a
generalized del Pezzo are present, and that no $-1$ curves not needed
for this blow down are included.  Such a network can be depicted
schematically as a graph where a set of nodes (the $-4$ curves) are
connected by edges (the $-1$ curves) (see
Figure~\ref{f:4-configurations}).
If we denote by $N$ the number of $-4$ curves in an allowed
configuration on a base associated with a theory having a given value
of $T$, $B$ the number of blow-downs that must be done to reach a
generalized del Pezzo, and $R$ the number of $-2$ curves remaining
after the blow-downs, we have from the bound on the number of $-2$
curves $R \leq 12$
\begin{equation}
\frac{3}{4} R + B \leq T \leq N + 9 \,.
\label{eq:4-bound1}
\end{equation}
We can, however, determine a lower bound on the ratio
\begin{equation}
\rho = \frac{3 R/4 + B}{N}  \geq \frac{23}{16}  \,.
\label{eq:4-bound}
\end{equation}
For any connected graph, this bound holds.  For example, a graph
formed from a loop of $n$ $-4$ curves connected by $-1$ curves
(Figure~\ref{f:4-configurations} (a)) has $b = r = n$ so $(3r/4 + b)/n
= 7/4 > 23/16$.  As another example, the graph formed by a single $-4$
curve connected by $-1$ curves to 3 other $-4$ curves
(Figure~\ref{f:4-configurations} (b)) has $b = 4, r =3, n = 4$, so
$(3r/4 + b)/n = 25/16 > 23/16$.  The bound is matched by a graph
component consisting of a $-4$ curve connected to 3 other $-4$ curves
of which one is terminal and two connect by a single edge to another
graph component --- this blows down to a $-2$ curve intersecting a
pair of $-1$ curves at a single point and has a contribution of $r =
1, b = 5, n = 4$ (counting the blow-down of each of the two
``external'' $-1$ edges as $b = 1/2$) for a total of $(3r/4 + b)/n =
23/16$.  The simplest example of a graph formed from a single
component of this type is shown in Figure~\ref{f:4-configurations}
(c).  This graph has $B = 5, R = 1, N = 4,$ so $\rho= 23/16$; note
that the blown-down configuration contains in addition to the $-2$
curve a pair of $-1$ curves not shown in the figure, which intersect
each other in two points, 
one of which is along the $-2$ curve.
A somewhat tedious case-by-case analysis of various
components shows that the bound (\ref{eq:4-bound}) cannot be exceeded.
(For example, extending any diagram by including an extra node on a
linear string of nodes contributes a factor of $7/4$ just as in the
closed loop example above).  From (\ref{eq:4-bound}) and
(\ref{eq:4-bound1}) it follows that $7 N/16 \leq 9 \Rightarrow N \leq
20$, from which it follows that $T \leq 29$ for any F-theory base
corresponding to a 6D supergravity theory with only $\gso(8)$ gauge
algebra summands.  This shows how in principle $T$ can be bounded for
a given class of gauge algebras.  In fact, this bound is weaker than
the true limit.  
The configurations that realize the bound
(\ref{eq:4-bound}) give $-2$ curve configurations that cannot appear
in a generalized del Pezzo with $T = 9$ \cite{Persson, Miranda}.  It
seems likely that the maximum $N, T$ that can be realized in practice is
$N = 12, T = 21$.
This can be achieved by a set of loops as in
Figure~\ref{f:4-configurations} (a)
of total length 12, or 3 copies of Figure~\ref{f:4-configurations}
(b).  
Note that a loop of this type of length 12 is not possible.  One
possible configuration of total length 12 comes from an allowed
configuration of $-2$ curves on  $dP_9$ consisting of nine $-2$
curves in a
closed loop and three $-2$ curves in a closed loop of length one ({\it
  i.e.}, with a single self-intersection).  It is not possible to
achieve a closed loop of alternating $-4, -1$ curves in the toric
context since this would blow down to a closed loop of $-2$ curves,
not a Hirzebruch surface.  The largest related toric base without
curves of self-intersection $-5$ or below has a loop of curves
containing six $-4$ curves and three $-3$ curves in 3 repeated copies of the
pattern $(-1, -4, -1, -3, -1, -4, \ldots)$.  Non-toric blow-ups of the
3 $-3$ curves give a loop containing  nine $-4$ curves with the structure
of Figure~\ref{f:4-configurations} (a).
Note also that while 4 copies of the diagram in
Figure~\ref{f:4-configurations} (b) does
not violate the bounds stated above, the detailed list of possible
$-2$ configurations  on a rational elliptic surface only
allows for three of the Kodaira type IV configurations of three
intersecting $-2$ curves.  In principle such constraints can be
included in a stricter bound that seems likely to rule out all
configurations with $N > 12$, but we do not pursue the details of such
an argument here.  This  discussion is intended only to give a
flavor of how one might construct a general argument bounding $T \leq
193$, where a much more complicated set of configurations involving
all the possible clusters from Figure~\ref{f:clusters}
must be
treated.  We leave further analysis along these lines for future work.

\subsection{Constraints on 6D supergravity theories}
\label{sec:constraints}

In ten dimensions, the macroscopic conditions of
supersymmetry and anomaly cancellation conditions
constrain supergravity theories so strongly that all spectra
compatible with these conditions are realized as string theory vacua
\cite{gs, heterotic, 10D}.  A theme in a series of recent works has been to investigate the extent to which a similar statement is true in
six dimensions by exploring the space of 6D theories compatible with
known constraints \cite{universality, finite, KMT, Park-Taylor}.
F-theory, in its current formulation, imposes additional constraints
on 6D supergravity theories; because of the close correspondence
between 6D supergravity data and F-theory geometry, these F-theory
constraints can be interpreted in a straightforward fashion as conditions on the spectrum
and action of 6D supergravity theories \cite{KMT-II}.  One constraint
imposed by F-theory on 6D supergravity is that the lattice of dyonic
string charges in the 6D theory be unimodular.  This turns out to be a
consistency constraint on 6D supergravity theories independent of
F-theory \cite{Seiberg-Taylor}, and therefore limits the
space of  theories.  Other constraints arise from F-theory for
which it is not yet known whether the constraints are actually
necessary for consistency of  quantum supergravity KEK in 6D, or
if the constraints are simply associated with F-theory and can be
avoided in another approach to quantum gravity, such as a more general
string compactification mechanism.  One such constraint on the
low-energy theory follows from the ``Kodaira condition'' in F-theory
stating that the total discriminant locus is $[\Delta] =-12K$
\cite{KMT-II}.  In the low-energy theory, $-K$ appears in the
gravitational Green-Schwarz coupling of the form $a \cdot B \wedge R
\wedge R$, where $K \rightarrow a$, and the divisor classes $D_i$ on
which 7-branes are wrapped to give nonabelian gauge groups are encoded
in the gauge Green-Schwarz couplings of the form $b_i \cdot B \wedge F
\wedge F$, where $D_i \rightarrow b_i$.  These terms are constrained
by the Kodaira condition since $-12K-\sum_{i} \nu_iD_i$ must be an
effective divisor (where $\nu_i$ is the multiplicity of 7-branes
wrapped on $D_i$) and therefore the corresponding combination of Green-Schwarz
coefficients must give a certain sign when dotted into the K\"ahler
modulus $j$, which is a vector under $SO(1, T)$ associated with the
scalars in the $T$ tensor multiplets.

In the analysis of \cite{clusters} and this paper, we used a key
feature of the algebraic geometry, which is that whenever there
exists an irreducible effective divisor $C$ with $C \cdot C \leq -3$,
it is necessarily the case that $-K \cdot C < 0$ and there must be a
nonabelian gauge group factor associated with this divisor.  This
condition has a corresponding interpretation in the resulting
low-energy supergravity theory.  For any low-energy 6D supergravity
theory arising from F-theory, if there is a supersymmetric dyonic
string state of charge $b$ such that the inner product
$b \cdot b$ associated with Dirac quantization satisfies $b \cdot b \leq
-3$, then $b$ must be the coefficient in a Green-Schwarz coupling $b
\cdot B \wedge F \wedge F$ for some gauge group factor containing at
least the minimal gauge algebra associated with that negative
self-intersection number through Table~\ref{t:clusters}.  This is true
as well for the combinations of self-intersection numbers $(-3, -2)$,
{\it etc.} giving non-Higgsable clusters containing multiple
divisors.  This condition is a highly nontrivial constraint on 6D
supergravity theories, relating the structure of the dyonic string
spectrum to the gauge and matter structure of the theory.  This
condition is also sufficient to imply the Kodaira condition.  It would
be very interesting to identify some reason in low-energy supergravity
why this kind of condition must hold.  If this could be done it would
place very strong additional constraints on the space of consistent 6D
supergravity theories and would be a major step forward towards
matching the set of consistent gravity theories with those that can be
realized from F-theory or other string compactifications.

The other key feature that was used in the analysis of this paper was
the pattern of divisor intersections produced by a blow-up of a point
on the base.  While the corresponding transitions between 6D
supergravity theories with different numbers of tensor multiplets have
been identified as tensionless string transitions \cite{dmw,
  Seiberg-Witten, Morrison-Vafa-II}, these transitions are not well
understood from the supergravity point of view.  Showing that
low-energy physics requires these transitions to impose changes in the dyonic string
lattice matching the changes in intersection structure used in this
paper would be another useful step in limiting the range of
possibilities for consistent 6D theories to match with what can be
produced from F-theory.

It is also possible that more general types of F-theory
compactifications may be possible, such as using orbifold bases or
incorporating additional structure
such as the ``T-branes'' of \cite{T-branes}.  It is possible that some
variations of F-theory or other string compactifications, such as on
asymmetric orbifolds, may give rise to low-energy theories that
violate the Kodaira constraint or other apparent constraints from
F-theory.  If it can be shown that string theory can indeed produce
additional 6D theories beyond those realized through conventional
F-theory, it would be desirable to understand how these fit into the
larger space of theories and match with low-energy constraints.

\subsection{4D F-theory models with toric bases}

There is a parallel structure between F-theory compactifications to
six dimensions and to four dimensions.  In both cases, there is an
underlying geometric moduli space of elliptically-fibered Calabi-Yau
manifolds that provides a substrate for understanding the space of
supersymmetric vacua.  In 6D this moduli space appears directly in the
low-energy theory, as all moduli are massless scalar fields.  In 4D,
this moduli space is obscured from the point of the low-energy theory
since many moduli are lifted by the necessary inclusion of fluxes, as
well as other perturbative and nonperturbative effects.  Nonetheless,
studying the moduli space of elliptic fibrations with section whose
total space is a Calabi-Yau manifold provides a clear mathematical
starting point for global studies of the space of 4D supersymmetric
F-theory vacua just as in 6D.  Just as the Green-Schwarz terms
appearing in 6D theories provide a natural connection to the geometry
of F-theory, similar axion-curvature squared terms that arise in 4D
theories characterize the geometry of an elliptically fibered fourfold
\cite{Grimm-Taylor}.  As the minimal model analysis provides a
systematic way of treating complex surfaces that can be bases for 6D
F-theory compactifications, the more general approach of Mori theory
\cite{Mori} gives a similar approach to treating complex threefold
bases.  Unlike in 6D, where a series of exceptional curves can be
blown down resulting in a minimal base that is either $\P^2$, a
Hirzebruch surface $\F_m$, or the Enriques surface, in 4D the set of
minimal bases is much larger and includes a variety of singular
spaces.  Nonetheless, in the toric context the general Mori program
simplifies and is clearly understood from a mathematical perspective.  So a
systematic analysis of toric F-theory bases for 4D supergravity
theories along similar lines to the work in this paper should be
possible.  
Previous analyses of a variety of toric constructions for 4D F-theory models
have been carried out in, for example, \cite{Mohri,Kreuzer-Skarke-4, kkmw}.
For threefold bases the story is complicated by the fact
that either points or curves can be blown up, and the curve structure
can be changed by ``flips'' and ``flops'' that leave the divisor
structure invariant.  While for 6D
models the tensionless string transition associated with blowing up a
point in the base involves trading 29 scalar fields for one tensor
field, the analogous geometric transition for 4D theories involves
blowing up a curve, with a more complicated change in the spectrum.  Another  kind of transition that is less well
understood from the physics perspective
involves blowing up a point to a divisor, requiring the
tuning of 481 scalar fields \cite{Klemm-lry, Grimm-Taylor}.
Just as for the
transitions involving blowing up points in a base surface this tuning
of moduli is visible in the Weierstrass monomials for a base toric
threefold.

The other key element in the analysis of this paper that must be
generalized systematically to 4D is the correspondence between the
intersection structure of divisors in the base and necessary structure
in the gauge group and matter content of the low-energy theory.  For
base surfaces this is relatively straightforward because the
intersection theory is simply governed by an integral lattice with
known properties.  For base threefolds, this is more complicated as
the intersection structure is described by a triple intersection form
on divisors.  Again, however, this structure simplifies for toric
bases.  We can systematically identify all local ``non-Higgsable
clusters'' for toric threefold bases in a similar fashion to the
analysis of \cite{clusters}.  To identify all possible structures for
a single divisor, we can locally consider a $\P^1$ bundle over 
a set of surfaces including each of
the 61,569 toric base surfaces identified in this paper.
Systematically identifying all possible local toric structures for
the section corresponding to a given surface $B$ will give all
single-component NHC's for a toric surface divisor of that topology.
We can then systematically combine these NHC's as in \cite{clusters},
but in a 2D triangulation, to get a list of all local non-Higgsable
clusters for 3D bases.

As a simple example of toric threefold bases for which minimal
gauge structure must be present, which also illustrate the simplest
NHC structure for threefold bases,
there is a family of toric threefolds
$\tf_m$ that are closely analogous to the Hirzebruch surfaces.  These
are $\P^1$ bundles over $\P^2$, with the single parameter $m$
characterizing the bundle.  F-theory on $\tf_m$ is dual to the
heterotic theory on a Calabi-Yau manifold given by an elliptic
fibration over $\P^2$ \cite{Berglund-Mayr, Grimm-Taylor}.  A toric
description of $\tf_m$ is given by a fan in which the 1D cones are
generated by
\begin{equation}
\tilde{F}_m: \;
v_1 =(1, 0, 0), 
v_2 =(0, 1, 0), 
v_3 =(-1, -1, m),
v_4 =(0, 0, 1), 
v_5 =(0, 0, -1)\,.
\end{equation}
The 2D cones for this model are generated by all pairs of $v_i, v_j, i
\neq j$ except $v_4, v_5$, and the 3D cones are given by all triples not
including both $v_4$ and $v_5$.  Just as for the Hirzebruch surfaces
$\F_m$ with $m \geq 3$, for the bases $\tf_m$  on the divisor $v_4$
$f$ and $g$ must vanish to sufficiently high degree for a nonabelian
gauge group through the Kodaira classification.  
To see this explicitly, using the same logic as in Section
\ref{sec:Weierstrass}, the lowest degree of vanishing $n$ of an allowed
section of $-aK$ on the divisor $D$ associated with $v_4$ (the section
of the $\P^1$ fibration) is realized at the point $x=(-a, -a, n -a)$
in the dual lattice $M$, which must satisfy $x \cdot v_3 = (2 -m)a +
mn\geq -a$, so the degree of vanishing of $f, g$ on $D$ must be at
least
\begin{equation}
[-aK] \geq  a(1-3/m)  \,, \;\;\;\;\; a = 4, 6 \,.
\label{eq:vanishing-degree}
\end{equation}
Thus, for example, on $\tf_4$ the degrees of vanishing of $f, g, \Delta$ are at
least 1, 2, 3, giving an $A_1$ singularity with a nonabelian gauge
algebra summand $\gsu(2)$.  Similar results hold for larger $m$, with
increasingly large gauge groups, up to $m = 18$ with an $\ge_8$ gauge
algebra, beyond which the singularity becomes too bad to support an
F-theory model; this matches with the dual heterotic theory where the
number of instantons in each $E_8$ factor of the gauge group is $18
\pm m$ \cite{Berglund-Mayr, Grimm-Taylor}.  This kind of analysis can
be carried out for other base threefolds, using a general form of the
Zariski decomposition (\ref{eq:Zariski}), though the mathematics is
more complicated when the base is non-toric.  The structure of the
$\tf_m$ around $v_4$ describes all possible local toric configurations
for a $\P^2$ divisor, so that the divisors up to $m = 18$ classify all
NHC's on a single $\P^2$ divisor.  As discussed above, this process
can be repeated for $\P^1$ bundles over all allowed
base surfaces to get a complete list of NHC's on a single divisor, and
iterated to find all NHC's for threefold bases.

We can in
principle explore the space of all possible toric threefold bases for
elliptic fibrations in a similar fashion to the analysis of this
paper.  By starting with a minimal set of bases, and then blowing up
and performing flips and flops in all ways compatible with the bound
on degrees of $f$ and $g$ on all divisors, curves, and points, we can
construct all toric threefold bases for elliptically fibered
Calabi-Yau fourfolds.  This will be a computationally more extensive
endeavor, however, then the classification of surface bases in this
paper.

It would also be interesting to systematically analyze the scalar
degrees of freedom in 4D F-theory models from the point of view of the
toric monomials.  Combining the singularity structure and Weierstrass
formulation for 4D theories should lead to a similar correspondence
between the 4D spectrum and 4D toric data; though there is no
gravitational anomaly in 4D that gives a condition analogous to
\eq{eq:constraint}, there is a close parallel between geometric
constraints on 4D theories from F-theory compactifications and the
underlying geometry \cite{Grimm-Taylor}.  Though all this structure is
less apparent in four dimensions than in six due to the lifting of
moduli as mentioned above, a systematic analysis of the part of the
theory dependent only on the underlying geometry should be possible.
We leave further analysis of these questions for future
work.

\section{Conclusions}
\label{sec:conclusions}

We have systematically classified all toric bases for elliptically
fibered Calabi-Yau threefolds.  These bases can be used to construct
6D supergravity theories from compactification of F-theory.  There are
61,539 such bases, giving supergravity models that typically have $T
\sim 25$ tensor multiplets, and have at most $T = 193$ tensor
multiplets.  We have some evidence and heuristic arguments that no
larger value of $T$ is possible even for non-toric bases, though we do
not have a rigorous proof of that statement.  For values of $T$ much
above 25, the gauge group has a rank that grows linearly in $T$ and is
dominated by gauge algebra summands $\ge_8,\gf_4$ and $(\gg_2
\oplus\gsu (2))$ with minimal non-Higgsable matter.

We have systematically determined the monomials in the Weierstrass
equation for these bases from the toric data and matched with the
degrees of freedom allowed from the gravitational anomaly constraint
(\ref{eq:constraint}).  We find that non-Higgsable matter fields do
not appear in the set of Weierstrass parameters.  In general, the
degrees of vanishing of $f, g$ on the curves of negative
self-intersection in the base are those associated with the minimal
(non-Higgsable) gauge and matter content described in \cite{clusters}.

Much of the analysis of this paper can be carried through in a similar
but more complicated context in four dimensions.  For four-dimensional
supergravity theories, the underlying geometry of the moduli space of
elliptically fibered fourfolds is expected to have an analogous
(though more complicated) structure to the space of elliptically
fibered threefolds.  The space of physical theories is less directly
related to this moduli space, however, as many moduli are lifted by
fluxes that must be included to satisfy tadpole cancellation
conditions.  We expect, however, a similar structure for 4D theories,
in which base manifolds of more complicated topology require
increasingly large gauge groups, corresponding to the non-Higgsable
gauge group factors appearing at large $T$ in the 6D theory.  It will
be interesting to understand whether in fact large non-Higgsable gauge
groups are in some sense generic for 4D theories.  There are many ways
in which this genericity may be avoided, however.  The larger number
of moduli for theories with smaller $T$ in the 6D context likely
corresponds to an exponentially larger number of discrete flux vacua
in the 4D context (see \cite{Douglas-Kachru, Denef-F-theory} for
reviews of flux vacua), so the more complicated bases may be
suppressed by this effect.  Codimension 3
loci where $f, g, \Delta$ vanish to orders at least $4, 6, 12$ 
necessarily involve surfaces as limiting fibers in the elliptic fibration
and -- unlike the codimension 2 case -- this cannot be cured by
blowing up the base \cite{codimthree}; it is unknown what the
physical interpretation of such 4D F-theory vacua might be.
Or there may be additional
effects that mitigate the large gauge groups expected for more
complicated 4D F-theory bases.  We leave further investigation of this
question to future work.

While we have primarily focused in this paper on the application to
F-theory, the set of bases for elliptically-fibered Calabi-Yau
threefolds that we have identified here have other potential
applications.  Understanding the set of elliptically-fibered
threefolds may shed light on the difficult problem of classifying
general Calabi-Yau threefolds.  Elliptically fibered Calabi-Yau
threefolds over the bases described here can also be used to construct
a very general class of heterotic string compactifications to 4D that
have F-theory duals.  As described in \cite{Vafa-f, Morrison-Vafa,
  Morrison-Vafa-II}, in general heterotic compactification to
dimension $10-2n$ on an
elliptically fibered Calabi-Yau $n$-fold over a base $B_{n-1}$
of complex dimension $n-1$
is dual to an F-theory compactification over a base $B_{n}$ that is a $\P^1$
fibration over $B_{n-1}$.  
Heterotic/F-theory
duality for 4D theories
has been studied for specific bundle constructions on the
heterotic side, for example in the stable degeneration limit \cite{fmw}.  A
general topological characterization of the duality independent of the
type of bundle construction follows from consideration of
axion-curvature squared couplings in 4D supergravity
\cite{Grimm-Taylor}.  
The bases given here give a broad range of
spaces in which this duality between 4D supergravity theories can be
studied in further detail.  

\vspace*{0.1in}

{\bf Acknowledgements}: We would like to thank Thomas Grimm, Vijay
Kumar, Joe Marsano, and
Daniel Park for helpful discussions.  Thanks to the the Aspen
Center for Physics for hospitality while part of
this work was carried out, and to the Simons Center for Geometry and
Physics, where this work was completed.
This research was supported by the DOE under contract
\#DE-FC02-94ER40818, and by the National Science Foundation under
grant DMS-1007414

\newpage

\appendix

\section{Appendix:  Systematic construction of all toric bases}

\begin{figure}
\begin{center}
\begin{picture}(200,160)(- 100,- 80)
\thicklines
\multiput(-130,50)( 0,-100){2}{\line(1, 0){60}}
\multiput(70,50)( 0,-100){2}{\line(1, 0){60}}
\put(-120,-60){\line( 0,1){120}}
\put(-70,60){\line( -1, -1){30}}
\put(-60,-30){\line( -1, -1){30}}
\put(90,60){\line( -1, -1){30}}
\put(100,-30){\line( -1, -1){30}}
\put(130,60){\line( -1, -1){30}}
\put(140,-30){\line( -1, -1){30}}
\thinlines
\put(-100,40){\line(1, -1){30}}
\put(-90,-10){\line(1, -1){30}}
\put(100,40){\line(1, -1){30}}
\put(110,-10){\line(1, -1){30}}
\put(60,40){\line(1, -1){30}}
\put(70,-10){\line(1, -1){30}}
\put(-80,0){\makebox(0,0){$\vdots$}}
\put(80,0){\makebox(0,0){$\vdots$}}
\put(120,0){\makebox(0,0){$\vdots$}}
\put(-140,0){\makebox(0,0){$0$}}
\put(-100, 60){\makebox(0,0){$m-k$}}
\put(-105, -60){\makebox(0,0){$-m$}}
\put(-80,43){\makebox(0,0){$-1$}}
\put(-80,27){\makebox(0,0){$-2$}}
\put(-86,-45){\makebox(0,0){$-1$}}
\put(-86,-25){\makebox(0,0){$-2$}}
\put(-50,5){\makebox(0,0){$k-1$ linked}}
\put(-50,-7){\makebox(0,0){$-2$ curves}}
\put(100, 63){\makebox(0,0){$m-k-l$}}
\put(95, -60){\makebox(0,0){$-m$}}
\multiput(81,43)(40,0){2}{\makebox(0,0){$-1$}}
\multiput(81,27)(40,0){2}{\makebox(0,0){$-2$}}
\multiput(73,-44)(40,0){2}{\makebox(0,0){$-1$}}
\multiput(73,-25)(40,0){2}{\makebox(0,0){$-2$}}
\put(50,7){\makebox(0,0){$l-1$ linked}}
\put(50,-5){\makebox(0,0){$-2$ curves}}
\put(150,5){\makebox(0,0){$k-1$ linked}}
\put(150,-7){\makebox(0,0){$-2$ curves}}
\end{picture}
\end{center}
\caption[x]{\footnotesize All toric bases for 6D F-theory models with
  $T > 1$ can be constructed by starting with one of these basic
  ``frames'' and blowing up only points at the intersection between
  divisors contained within one of the vertical fibers.  The adjacent
  $(-1, -2)$ and $( -2, -2)$ curves can be independently blown up into
  a relatively small set (485 and 207 possibilities respectively) of
  possible links that can be attached into chains using a recursive
  algorithm that efficiently enumerates all possibilities.}
\label{f:skeletons}
\end{figure}
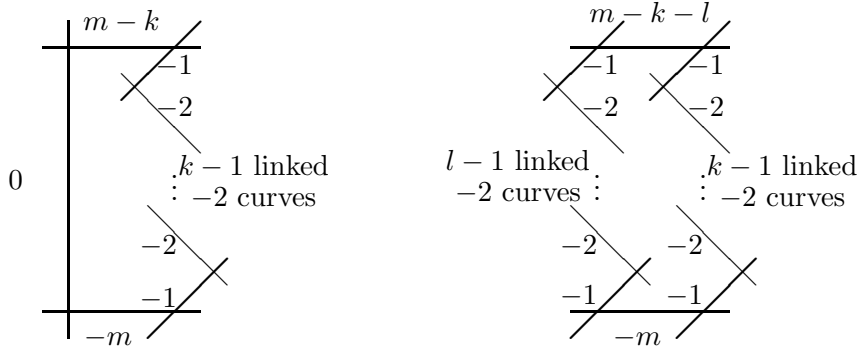

In this appendix we give a brief description of the algorithm used for
systematically constructing all toric bases.  The basic idea is to
start with the Hirzebruch surfaces and blow up intersection points
between divisors until no further blow-ups are possible.  Because this
leads to a large  number of
combinatorial possibilities, it is helpful
to structure the tree of possibilities in a compact way.  
The Hirzebruch surfaces $\F_m$ have four intersecting divisors.  There
is a section $S_\infty$ of self-intersection $S_\infty \cdot S_\infty = -m$, and another
section $S_0= S_\infty+ mF$ of self-intersection $S_0\cdot
S_0= m$.  There are two fibers $F$ that are equivalent
homologically, that can be pictured as vertical curves each of which
intersects both $S_\infty$ and $S_0$.  For notational convenience we
call these $F_1$ and $F_2$.  We organize the possible
results of a sequence of
blow-ups by starting with the blow-ups at the points where
$S_0$ intersects the fibers $F_i$.  If we start on $\F_m$ and
blow up the intersection of $S_0$ with $F_1$, then the self-intersection of
$S_0$ goes down by  one, and the fiber $F_1$ is replaced by a
pair of $-1$ curves that intersect one another, one intersecting
$S_0$ and the other intersecting $S_\infty$.  Blowing up again at the
intersection with $S_0$, the fiber becomes a sequence of curves
$(-1, -2, -1)$.  Repeating for a total of $k$ blow-ups, the
self-intersection of $S_0$ reduces by  $k$, and the fiber $F_1$
is replaced by a sequence $(-1, -2^{k-1}, -1)$, with $k-1$ $-2$ curves
between two $-1$ curves.  We can then do the same thing with fiber
$F_2$.  Blowing up $l$ times at the intersection with $S_0$ we
end up with a toric base described by a sequence of curves of
self-intersection (see Figure~\ref{f:skeletons})
\begin{equation}
I = ( m-k-l, -1, -2^{l-1}, -1, -m, -1, -2^{k-1}, -1) \,.
\label{eq:skeletons}
\end{equation}
Starting with these ``frames,'' for all possible $m, k, l$ all
possible toric bases can be realized by blowing up points between
pairs of curves in the chains replacing the fibers $F_i$.  Note that
if a point at the intersection of $S_\infty$ with either fiber $F_i$ is blown
up, it gives an identical chain, though with an increase by one in
$m$.  Similarly, if $m-k-l < -m$, then the resulting chain can be
realized starting from $\F_{k + l-m}$, with roles of the two divisors $S_\infty, S_0$
reversed.  Thus, we consider all possible ``frames'' of the form
\eq{eq:skeletons}, with $k + l \leq 2m, m \leq 12$.  

We can
consider all blow-ups of the chains on each side of the ``frames''
\eq{eq:skeletons} by considering a
sequence of blow-ups of the pairs $(-1, -2)$ and $(-2, -2)$.  For
example, blowing up $(-1, -2)$ gives $(-2, -1, -3)$; blowing this up
again gives either $(-3, -1, -2, -3)$ or $(-2, -2, -1, -4),$ {\it
  etc.}.  Continuing this leads to 207 distinct ways in which the pair
$(-2, -2)$ can be blown up to a sequence that satisfies the F-theory
rules contained in Table~\ref{t:clusters} and
Table~\ref{t:connections}.  For $(-1, -2)$ there are 485
possibilities.  Note that in some cases, intermediate stages in the
blow-up process do not satisfy the rules of Table~\ref{t:connections};
any intersection between clusters that are so large that $\Delta$ has
degree 12 or greater at the intersection point must be blown up until
the base is acceptable.  If any divisor develops a self-intersection
of $-13$ or below then no further valid bases can be reached from this
point.  Given the ways of blowing up the $(-2, -2)$ pairs, we can
attach these ``links'' into chains recursively, by adding the change
in self-intersections on the end divisors from adjacent pairs.  For
example, a pair $(-2, -2)$ blows up to $(-3, -1, -3)$.  Connecting a
pair of these in a chain gives $(-3, -1, -4, -1, -3)$.  In many cases,
the decrease in self-intersection at the endpoints makes it impossible
to connect two links.  This dramatically decreases the combinatorial
growth of the ways that links can be attached.  For example, attaching
the $k$ links arising from blowing up each of the pairs in the
sequence $(-1, -2^{k-1}, -1)$ gives rise to just 1107 possibilities
for any number $k \geq 12$.  Given the various possibilities for
attaching links we can construct the possibilities for each blown-up
fiber $F_i$, and attach these to the sections $S_\infty, S_0$ in all
possible ways compatible with the F-theory rules in Table~\ref{t:clusters} and
Table~\ref{t:connections}.  The combinatorics of this algorithm are
fairly straightforward and the complete calculation can be done with a
few hours of computer time in Mathematica or a similar higher-level
computational package.  Note that in the case $k = 1$ or $l = 1$ a
fiber is replaced by only a pair of $-1$ curves.  In this case we can
proceed by blowing up the intersection point giving a sequence $(-2,
-1, -2)$.  If we then continue by blowing up the intersection with the
first curve another $r -1$ times we get the sequence $(-(1 + r), -1,
-2^r)$, and we can proceed as before by blowing up the $(-1, -2)$ and
$(-2, -2)$ pairs to get links that we attach recursively.

This gives a systematic algorithm for constructing all toric bases
that we have implemented to give the results described in Section
\ref{sec:toric}.  Note that this algorithm will produce some bases in
multiple inequivalent ways.  After expanding all possible ``frames''
(\ref{eq:skeletons}) in all possible ways, we must put all resulting
bases in a canonical form (such as decreasing ``dictionary'' ordering
by rotating (and possibly reflecting) the chain so that the largest
self-intersection is at the initial position etc.)  Taking only one
base in each canonical form gives a list in which each of the 61,539
distinct toric bases appears precisely once.  Essentially the same
algorithm can be used to construct a wide family of non-toric bases,
where we consider more than two fibers $F_i$ and only blow up points
that are either at the intersection of an $F_i$ with $S_\infty$ or
$S_0$, or at the intersection of two curves within a single
$F_i$.  Some partial results on these more general families of bases
are described briefly in Section \ref{sec:non-toric}.

A complete listing of the 61,539 distinct toric bases can be found
online at \cite{data}.

\end{document}